\documentclass[12pt,a4paper]{article}
\pdfoutput=1
\usepackage{jheppub}
\usepackage{amsfonts}
\usepackage{bbm}
\usepackage{multirow}
\usepackage{enumitem}

\def \Z{\mathbb{Z}}
\def \R{\mathbb{R}}
\def \C{\mathbb{C}}
\def \Q{\mathbb{Q}}
\def \CL{\mathcal{L}}
\def \CN{\mathcal{N}}
\def \lk{\ell k\,}
\def \vert{V}

\title{Supergroups, q-series and 3-manifolds}

\author[1,2,3]{Francesca Ferrari,}
\author[3]{Pavel Putrov}

\affiliation[1]{SISSA, Via Bonomea 265, Trieste 34136, Italy }
\affiliation[2]{INFN, Sezione di Trieste, Via A. Valerio 2, Trieste 34127, Italy}
\affiliation[3]{ICTP,
Strada Costiera 11, Trieste 34151, Italy}

\abstract{We introduce supergroup analogues of 3-manifold invariants $\widehat{Z}$, also known as homological blocks, which were previously considered for ordinary compact semisimple Lie groups. 
We focus on superunitary groups, and work out the case of $SU(2|1)$ in details. Physically these invariants are realized as the index of BPS states of a system of intersecting fivebranes wrapping a 3-manifold in M-theory.
As in the original case, the homological blocks are $q$-series with integer coefficients. 
We provide an explicit algorithm to calculate these $q$-series for a class of plumbed 3-manifolds and study quantum modularity and resurgence properties for some particular 3-manifolds. Finally, we conjecture a formula relating the $\widehat{Z}$ invariants and the quantum invariants constructed from a non-semisimple category of representation of the unrolled version of a quantum supergroup.}

\begin{document}
\maketitle

\section{Introduction}

Topological quantum field theories (TQFTs) play an important role in description of infrared dynamics of physical theories. They also provide a useful tool to study the topology of manifolds. One of the most non-trivial known topological invariants of 3-manifolds and links is provided by $SU(2)$ Chern-Simons topological quantum field theory \cite{Witten:1988hf,reshetikhin1990ribbon,reshetikhin1991invariants}. In the mathematics literature the corresponding invariant of 3-manifolds is known as Witten-Reshetikhin-Turaev (WRT) invariant, while the corresponding invariant of links is the colored Jones  polynomial (or more generally HOMFLY-PT, corresponding to $SU(N)$ gauge group in Chern-Simons theory). The invariants of links and 3-manifolds are related to each other by the surgery construction. In \cite{khovanov2000categorification,khovanov2005categorifications,khovanov2008matrix} new homological invariants of links, that categorify the colored Jones polynomial,  were found. Their physics realization in terms of string/M-theory and also a 4d TQFT was provided in \cite{Gukov:2004hz,Witten:2011zz}. This physical realization also predicts the existence of the corresponding homological invariant of non-trivial 3-manifolds. However, one encounters many problems in an attempt to rigorously define such new 3-manifold invariants, or even calculate them. Some progress in this direction was made in \cite{Gukov:2016gkn,Gukov:2017kmk} where, motivated by physics, certain new invariants of 3-manifolds were considered, often referred to as \textit{homological blocks} or $\widehat{Z}$ (``\textit{Z-hat}'' invariants). For a given closed 3-manifold their value is a vector of $q$-series with integer coefficients and thus, similarly to the colored Jones polynomial of links, allows a categorification (unlike the WRT invariant itself, which a priori does not naturally contain any integer valued invariants). Physically the invariants $\widehat{Z}$ can be understood as the half-indices \cite{Gadde:2013wq,Yoshida:2014ssa} (i.e. the partition function on $D^2\times S^1$, cf. also \cite{Dimofte:2010tz,Beem:2012mb}) of a 3d $\mathcal{N}=2$ supersymmetric theory $T_{\mathfrak{sl}(2)}[M^3]$. The theory $T_{\mathfrak{sl}(2)}[M^3]$ is the effective 3d theory obtained by (topologically twisted) compactification of the 6d $\mathcal{N}=(2,0)$ superconformal theory of type $A_1$. In \cite{Gukov:2016gkn,Gukov:2017kmk} a general relation between the $\widehat{Z}$ invariants and WRT invariant was conjectured. 
From now on, we will denote these invariants as $\widehat{Z}^{\mathfrak{sl}(2)}$, to emphasize that they are related to Chern-Simons theory with gauge group $SU(2)$, with the corresponding Lie algebra being\footnote{We are considering Lie algebras over complex numbers throughout the paper, and thus there is no difference between $\mathfrak{sl}(n)$ and $\mathfrak{su}(n)$ for us.} $\mathfrak{sl}(2)$. This construction has a natural generalization to an arbitrary reductive Lie algebra\footnote{We choose to label $\widehat{Z}^{\mathfrak{g}}$ by a Lie algebra, instead of the corresponding gauge group in Chern-Simons theory, because of its 6d origin. As reviewed later, in the case when $\mathfrak{g}$ is of ADE type, or $\mathfrak{gl}(1)$, $\widehat{Z}^{\mathfrak{g}}$ invariant of a 3-manifold $M^3$ can be realized in terms of a 6d $\mathcal{N}=(2,0)$ superconformal field theory of type $\mathfrak{g}$ compactified on $M^3$. It is known that such 6d theories are specified by a choice of a Lie algebra, not a Lie group. In principle $\widehat{Z}^{\mathfrak{g}}$ can be related to Chern-Simons theories with different compact gauge groups $G$ that have $\mathfrak{g}$ as their Lie algebra. However, for simplicity of the discussion, in this work we will assume that $G$ is fixed to be simply-connected.} $\mathfrak{g}$, with the corresponding invariant being $\widehat{Z}^{\mathfrak{g}}$. In \cite{Gukov:2017kmk,Gukov:2019mnk,Park:2019xey,Chun:2019mal} an explicit algorithm to calculate $\widehat{Z}^{\mathfrak{g}}$ for a large class of 3-manifolds was given. However, at the moment there is no explicit proposal for the underlying homological invariants, except in the simplest examples when the 3-manifold is a lens space.

The homological invariants categorifying $\widehat{Z}^{\mathfrak{sl}(2)}$ are expected to be somewhat similar to Monopole Floer homology of 3-manifolds  \cite{kronheimer2007monopoles}. The latter homological invariant does have a mathematically rigorous definition: it is  known to be equivalent to  the later developed Heegard Floer homology of \cite{ozsvath2004holomorphic} and to be closely related to  Instanton Floer homology \cite{floer1988instanton}. In physical terms, Monopole Floer homology can be understood (up to certain subtleties) as the Hilbert space of the 4d Seiberg-Witten TQFT \cite{Witten:1994cg}, the topologically twisted $\mathcal{N}=2$ supersymmetric $U(1)$ gauge theory with a single charge one hypermultiplet. The decategorification  of Monopole Floer homology gives 3d Seiberg-Witten invariants of 3-manifolds, which are known to coincide with Reidemeister-Turaev-Milnor torsion \cite{meng1996underline}. The latter can also be realized by 3d Chern-Simons TQFT with gauge group being $U(1|1)$ supergroup \cite{Rozansky:1992zt}. The corresponding invariant of knots is the Alexander polynomial, which is categorified by knot Floer homology \cite{ozsvath2004holomorphic}. A string/M-theoretic realization of 3d Seiberg-Witten invariants and its categorification, Monopole Floer homology, was considered in \cite{Mikhaylov:2015nsa,Gukov:2016gkn}. In particular, the 3d Seiberg-Witten invariants can be understood as $\widehat{Z}^{\mathfrak{gl}(1|1)}$, the $\mathfrak{gl}(1|1)$ super Lie algebra version of the $\widehat{Z}^{\mathfrak{sl}(2)}$ homological blocks. The correspondence between various invariants associated to $SU(2)$ and $U(1|1)$ gauge groups in  Chern-Simons theory is summarized in  Table \ref{tab:invariants}.

\begin{table}
\footnotesize
\centering
	\begin{tabular}{|p{1.2cm}||p{1.8cm}|p{2.5cm}||p{1.8cm}|p{2.5cm}||p{2.2cm}|}
	\hline
	\multirow{2}{*}{Group} & \multicolumn{2}{|c||}{Knots} & \multicolumn{2}{|c||}{3-manifolds} & 4-manifolds \\
	\cline{2-6}
	& invariant & categorification & invariant & categorification & invariant \\
	\hline
	\hline
$U(1|1)$ & Alexander \newline polynomial
	 & knot Floer  \newline homology & 3d SW \newline invariants & Monopole Floer \newline homology & 4d SW \newline invariants \\
	 \hline
	$SU(2)$ & Jones \newline polynomial
	  & Khovanov  \newline homology & WRT \newline invariant & ? & ??? \\
	   \hline
	\end{tabular}
	\caption{\label{tab:invariants} The correspondence between various topological invariants associated to $SU(2)$ and $U(1|1)$ gauge groups in 3d Chern-Simons theory.}
\end{table}

This suggests that to understand the categorification of $\widehat{Z}^{\mathfrak{sl}(2)}$ it might be instructive to consider the version of $\widehat{Z}^{\mathfrak{g}}$ for more general super Lie algebras $\mathfrak{g}$, given that the categorification of $\widehat{Z}^{\mathfrak{gl}(1|1)}$ is already known. Motivated by this, in this work we develop a basic theory of the supergroup version of homological blocks. We in particular focus on the case of $\mathfrak{g}=\mathfrak{sl}(N|M)$ or $\mathfrak{gl}(N|M)$ (when the gauge group of the corresponding Chern-Simons theory is $G=SU(N|M)$ or $U(N|M)$ respectively) and work out many technical details for $G=SU(2|1)$. 
This is the simplest example that provides a connection between the $G=SU(2)$ and $G=U(1|1)$ cases in the Table \ref{tab:invariants} since
\begin{equation}
	U(1|1) \subset SU(2|1) \supset SU(2).
\end{equation}

The structure of the paper is as follows. In Section \ref{sec:brane-setup} we provide a brane realization of $\widehat{Z}^{\mathfrak{gl}(N|M)}$ (and their $\mathfrak{sl}$ and $\mathfrak{psl}$ versions), mostly following \cite{Mikhaylov:2014aoa}. In Section \ref{sec:Zhat-plumbed-general} we give a contour integral expression for $\widehat{Z}^{\mathfrak{gl}(N|M)}$  (without specifying the choice of the contour) in the case of plumbed 3-manifolds, which are reviewed in Section \ref{sec:plumbed-3-manifolds}.  In Section \ref{sec:Zhat-sl21} we fix the contour ambiguity in the case $\mathfrak{g}=\mathfrak{sl}(2|1)$ and consider explicitly the resulting $q$-series for some simple examples of 3-manifolds. A mathematically oriented reader can focus on equation (\ref{Zhat-plumbed-sl21-CT}) that gives a precise definition of the topological invariants $\widehat{Z}^{\mathfrak{sl}(2|1)}$ on a certain class of plumbed 3-manifolds. In Section \ref{sec:resurgence} we study resurgence properties of these invariants. In Section \ref{sec:q-mod} we analyze quantum modular properties of $\mathfrak{sl}(2|1)$ homological blocks of lens spaces. In Section \ref{sec:quantum-supergroup} we provide a conjectural relation between $\widehat{Z}^{\mathfrak{sl}(2|1)}$ and the invariants associated to a non-semisimple category of representations of the unrolled quantum supergroup $\mathcal{U}_q^H(\mathfrak{sl}(2|1))$ \cite{ha2018topological}.
Finally, in Section \ref{sec:open-questions} we provide a list of open questions. The appendices contain various technical arguments and calculations, the results of which are used in the main text.

\section{BPS states of intersecting fivebranes wrapping a 3-manifold}

\subsection{Brane setup}
\label{sec:brane-setup}

From the point of view of 11-dimensional M-theory we are interested in a setup containing two stacks of M5-branes. The ambient M-theory space-time and the supports of fivebranes are the following:
 	\begin{equation}
		\begin{array}{cccccccc}
				\text{M-theory} & T^*M^3&\times & \C&\times & \C &\times & S^1_\text{time} \\ 
				\text{$N$ M5-branes} & M^3&\times & \C&\times & \{0\} &\times & S^1_\text{time} \\ 
				\text{$M$ M5-branes} & M^3&\times &\{0\}&\times & \C &\times & S^1_\text{time} \\ 
			\end{array}
			\label{M-setup}
		\end{equation}		
where $M^3$ is a closed 3-manifold and $T^*M^3$ is the total space of its cotangent bundle. The factor $\C^2$ in the M-theory spacetime, as a Riemannian manifold, is the Taub-NUT space. It is usually represented as a circle fibration over $\mathbb{R}^3$ with a single vanishing fiber at the center (corresponding to a single Kaluza-Klein monopole). The $U(1)$ action on the circle fibers is $(z,w)\mapsto (e^{i\phi}z,e^{-i\phi}w)$, for $(z,w)\in \C^2$ and $e^{i\phi}\in U(1)$. We denote the corresponding groups as $U(1)_q$ and its generator as $L_0$. The two stacks of M5-branes (containing $N$ and $M$ branes) are supported on $z=0$ and $w=0$ subspaces of $\mathbb{C}^2$. They look like ``cigars'' in the Tab-NUT metric as they are embedded in the circle fibration over the $\R^3$ base as the fibration restricted to two rays originating at the origin. Note that in $\C^2$ these two cigars intersect transversally at a single point, while the corresponding rays in the $\R^3$ base go in the opposite directions from the origin. 

In the whole M-theory spacetime the stacks of fivebranes intersect transversally along $M^3\times S^1_\text{time}$, where $M^3$ is considered as the zero-section of the $T^*M^3$ bundle. This is a rather standard M-theoretic realization of topological string theory on the Calabi-Yau threefold $T^*M^3$ with $N$ branes and $M$ anti-branes wrapping its Lagrangian cycle $M^3$, argued to be dual to Chern-Simons theory on $M^3$ with the gauge group $U(N|M)$ \cite{Vafa:2001qf,Aganagic:2003db,Gorsky:2013jxa,Mikhaylov:2014aoa,Kozcaz:2018ndf}. 
It is also a generalization of the M-theory setup with $M=0$ considered in the study of $SU(N)$ and $U(N)$ homological blocks of $M^3$ in \cite{Gukov:2016gkn,Gukov:2017kmk}, and the setup with $N=M=1$ considered for $SU(1|1)$ homological blocks in \cite{Gukov:2016gkn,Dedushenko:2017tdw}. 

 Such configuration of fivebranes preserves 2 supercharges for a general 3-manifold $M^3$. The unbroken supercharges also commute with $U(1)_q$ symmetry. Therefore one can consider the corresponding flavored Witten index with $S^1_\text{time}$ treated as the time circle:
 \begin{equation}
     \widehat{Z}^{\mathfrak{gl}(M|N)}_{a,b}[M^3]:=
     \mathrm{Tr}_{\mathcal{H}_{a,b}} (-1)^F q^{L_0}
     \label{Zhat-index}
 \end{equation}
 where $\mathcal{H}_{a,b}$ denotes the BPS subspace (equivalently, the $Q$-cohomology, where $Q$ is a preserved complex supercharge satisfying $Q^2=0$) of the Hilbert space of the fivebrane system described above.
 The pair of indices $a,b$ denotes choices of certain boundary conditions at the ends of the two cigars (which are disks topologically). From this point on, to simplify our discussion, we will assume that the 3-manifold $M^3$ is a rational homology sphere, that is its first Betti number is zero: $b_1=0$. In \cite{Gukov:2017kmk}, where a single stack of $N$ fivebranes was considered, it was argued that there is a natural set of boundary conditions labeled by the elements of\footnote{In the case of $SU(2)$ group, it was later argued in \cite{Gukov:2019mnk,Gukov:2020lqm} that the more natural set of labels is the set of spin$^c$ structures on $M^3$, which is a torsor over $H^2(M^3,\Z)\cong H_1(M^3,\Z)$. We will address this subtlety later in the paper. Moreover, one can also assume that $M^3$ has a fixed spin-structure (any closed oriented 3-manifold is spin), which then can be used to fix canonically an isomorphism between the set of spin$^c$ structures and $H_1(M^3,\Z)$. } $H_1(M^3,\Z)^{N}$ (assuming the center of mass is not removed), modulo the action of the Weyl group of $U(N)$. Therefore in the case of two stacks with $N$ and $M$ branes there is a natural choice of labels $(a,b)\in H_1(M^3,\Z)^N\times H_1(M^3,\Z)^M$ modulo the action of the Weyl group of $U(N)\times U(M)$. The BPS states contributing to the index can be realized by M2-branes ending on the fivebranes. Namely, they can end on a 2-dimensional cycle of the form $\gamma   \times \{0,0\} \times S^1_\text{time}$ inside the worldvolume of a single fivebrane, where $\gamma\in M^3$ is a one dimensional cycle in $M^3$. Such M2-branes can be separated into charge sectors corresponding to the class $[\gamma]\in H_1(M^3,\Z)$. The choice of a pair $(a,b)$ in \eqref{Zhat-index} therefore can be understood as restriction to the sector $\mathcal{H}_{a,b}$ in the BPS spectrum with a fixed total charge.

 The fugacity $q$ in the trace $(\ref{Zhat-index})$ is related to the coupling constant $g_s$ of the topological string theory on $T^*M^3$ as $q=\exp{g_s}$. The topological string coupling constant is, in turn, related to the (analytically continued) coupling $k$ of Chern-Simons theory on $M^3$ as $g_s=\frac{2\pi i}{k}$. Moreover, the fugacity $q$ has the following geometric interpretation. The insertion of $q^{L_0}$ in the trace in (\ref{Zhat-index}) corresponds to twisting the metric on the product $\mathbb{C}^2\times S^1_\text{time}$. The boundary circles of the cigars at infinity, multiplied by $S^1_\text{time}$, form two copies of a 2-torus with complex structure $\tau$, such that $q=e^{2\pi i\tau}$ and $\tau \in \mathcal{H}$, the upper half of the complex plane. That is $\tau=1/k$ in terms of Chern-Simons level.
 
As explained in detail in \cite{Mikhaylov:2014aoa} (generalizing the argument for $U(N)$ Chern-Simons gauge group \cite{Witten:2011zz}), the relation of this system of fivebranes to the $U(N|M)$ Chern-Simons theory can be also understood through the following sequence of dualities. First, one can reduce M-theory to type IIA string theory by choosing the M-theory circle to be the fiber of the Taub-NUT space, considered as the circle fibration over $\mathbb{R}^3$. The M-theory setup (\ref{M-setup}) then translates into the following setup in type IIA string theory:
 	\begin{equation}
			\begin{array}{cccccc}
				\text{type IIA} & T^*M^3&\times & \R^3 &\times & S^1_\text{time} \\ 
				\text{$N$ D4-branes} & M^3&\times & \R_- &\times & S^1_\text{time} \\ 
				\text{$M$ D4-branes} & M^3&\times & \R_+ &\times & S^1_\text{time} \\ 
				\text{1 D6-brane} & T^*M^3&\times & \{0\} &\times & S^1_\text{time} \\ 
			\end{array}
			\label{type-IIA-setup}
		\end{equation}		
where $\R_\pm$ are positive and negative half-axes of a one-dimensional subspace $\R\subset \R^3$ passing through the origin.

Next one can perform T-duality along the $S^1_\text{time}$ circle to obtain the following setup in type IIB sting theory:
 	\begin{equation}
		\begin{array}{cccccc}
				\text{type IIB} & T^*M^3&\times & \R^3 &\times & S^1 \\ 
				\text{$N$ D3-branes} & M^3&\times & \R_- &\times &  \text{pt} \\ 
				\text{$M$ D3-branes} & M^3&\times & \R_+ &\times & \text{pt} \\ 
				\text{1 D5-brane} & T^*M^3&\times & \{0\} &\times & \text{pt} \\ 
			\end{array}
			\label{type-IIB-setup}
		\end{equation}
The worldvolume theories on the stacks of $M$ and $N$ $D3$ branes are $\mathcal{N}=4$ super-Yang-Mills (SYM) theories with gauge groups $U(M)$ and $U(N)$ respectively. Both theories have the same complexified gauge coupling $\frac{4\pi i}{g^2}+\frac{\theta}{2\pi}=\tau$. The D5-brane provides an interface between them imposing Nahm-pole-like boundary conditions on both sides \cite{Gaiotto:2008sa}. The theories are topologically twisted along $M^3\subset T^*M^3$.

\begin{figure}[tbp]
\centering 
\includegraphics[scale=1]{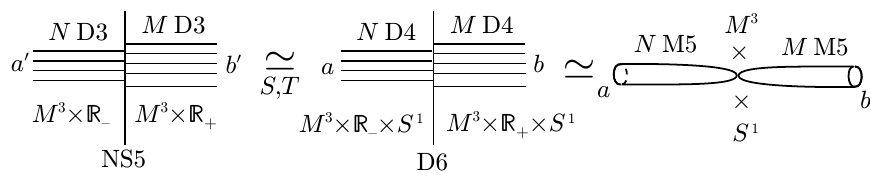}
\caption{A sequence of dualities relating M-theory and string theory realizations of $U(N|M)$ supergroup Chern-Simons theory on $M^3$.}
\label{fig:super-CS-branes}
\end{figure}

Finally, one can perform S-duality to replace the D5-brane with an NS5-brane. This also changes the coupling constant of the 4d SYM theories from $\tau$ to $-1/\tau=-k$. As was argued in \cite{Gaiotto:2008sd,Kapustin:2009cd,Mikhaylov:2014aoa}, the supersymmetry protected sector of the 4d SYM theories of $M^3\times \R_\pm$, coupled through the corresponding interface, can be described in terms of $U(N|M)$ supergroup Chern-Simons theory with level $k$. In particular the total action of the 4d-3d coupled system is equal to the action of the supergroup Chern-Simons theory on $M^3$ up to $Q$-exact terms. Note that in this setup it is natural to require that the gauge fields  in the 4d theories approach flat connections at the infinite ends of $\R_\pm$. A choice of such boundary conditions at two infinities can be understood as a choice of a flat connection of $U(N)\times U(M)$ on $M^3$. This group is the maximal bosonic subgroup of $U(N|M)$ gauge group of the Chern-Simons theory on $M^3$. In terms of the path integral of the Chern-Simons theory, the choice of such boundary condition corresponds to picking the contribution of the corresponding critical point. These boundary conditions however transform non-trivially under S-duality. The boundary conditions $(a,b)$ at the infinite ends of cigars in Figure \ref{fig:super-CS-branes} are in general non-trivial superpositions of the boundary conditions corresponding to flat connections in the Chern-Simons theory.

The sequence of dualities described above is schematically depicted in Figure \ref{fig:super-CS-branes}.  One can also consider the $\mathfrak{sl}(M|N)$ version of (\ref{Zhat-index}) corresponding to $SU(M|N)$ Chern-Simons theory. In the brane picture, this is realized by excluding the degrees of freedom corresponding to simultaneous translation of all the D3 branes in the normal direction (vertical direction in Figure \ref{fig:super-CS-branes}). When $N=M$ and $SU(N|N)$ has a $U(1)$ center subgroup, one can also consider the $\mathfrak{psl}(N|N)$ version of (\ref{Zhat-index}) corresponding to $PSU(N|N)$ gauge supergroup. In the brane picture this is realized by requiring that the positions of the centers of mass in the normal direction are the same for both stacks.

Finally, let us note that unlike in the case of ordinary Lie groups, the index \eqref{Zhat-index} (and its $\mathfrak{sl}$ and $\mathfrak{psl}$ versions) generically cannot be interpreted directly as a partition function on $D^2\times S^1_\text{time}$ of a 3d $\mathcal{N}=2$ theory $T_\mathfrak{g}[M^3]$ associated to the 3-manifold $M^3$ by the 3d/3d correspondence \cite{Dimofte:2011ju,Chung:2014qpa}. As in $\mathfrak{g}=\mathfrak{sl}(2)$ case, $T_\mathfrak{g}[M^3]$ denotes an effective 3-dimensional quantum field theory obtained by a twisted compactification a 6d $\mathcal{N}=(2,0)$ theory corresponding to a Lie algebra $\mathfrak{g}$. On the other hand, (\ref{Zhat-index}) can be interpreted as the partition function of a pair of quantum field theories living on intersecting spacetimes (similar setups have been considered in \cite{Gomis:2016ljm,Pan:2016fbl,Nieri:2017ntx,Nieri:2018pev}). Namely, consider first two \textit{different} 3d QFTs, $T_{\mathfrak{gl}(N)}[M^3]$ and $T_{\mathfrak{gl}(M)}[M^3]$ living on two \textit{different} copies of $D^2\times S^1_\text{time}$ (with opposite orientation). Identify then the one-dimensional subspaces $\{0\}\times S^1_\text{time}$ in both spacetimes and introduce certain 1d dimensional degrees of freedom (i.e. a quantum mechanics) supported on this common 1d subspace. These 1d degrees of freedom couple to both 3d QFTs. In the string theory picture they originate from strings stretched between two stacks of branes.

We will return to this point in Section \ref{sec:plumbed-3-manifolds} where we consider the case when $M^3=L(p,1)$, a lens space, and $T_{\mathfrak{gl}(N)}[L(p,1)]$ has an explicit rather simple Lagrangian description.

\subsection{Basic facts about plumbed 3-manifolds}
\label{sec:plumbed-3-manifolds}

We want to propose an explicit expression for the indices (\ref{Zhat-index}) for a certain class of 3-manifolds. Namely, consider a 3-manifold $M^3$ associated to a plumbing graph $\Gamma$. The latter is a simple weighted graph which consists of a set of edges, a set of vertices (which we denote by $V$) and a weight function which assigns to each vertex $I\in V$ an integral coefficient $a_I$. 

\begin{figure}[tbp]
\centering 
\includegraphics[scale=2.3]{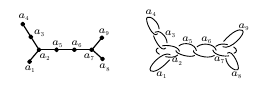}
\caption{An example of a plumbing graph $\Gamma$ and a corresponding framed link $\mathcal{L}(\Gamma)$ in a three-sphere.}
\label{fig:plumbing}
\end{figure}

There are various (equivalent) ways to define the 3-manifold corresponding to $\Gamma$. For instance, on can start with a framed link $\CL(\Gamma)\in S^3$ associated to the plumbing graph $\Gamma$ as illustrated in the example in Figure \ref{fig:plumbing}. That is, for each vertex $I$ we associate an unknot with framing specified by $a_I\in \Z$, its self-linking number. A presence of an edge between two vertices in $\Gamma$ corresponds to the fact that the pair of unknots associated to the two vertices is linked in a most standard way, namely the unknots form a Hopf link together. The 3-manifold $M^3$ is then obtained by Dehn surgery on the framed link $\CL(\Gamma)$. That is one removes tubular neighborhoods (which are isomorphic to solid tori) of all link components and glues them back after swapping the meridians and the longitudes on the boundary tori\footnote{As a reminder, a \textit{meridian} is a 1-cycle going once around the link, contractible inside the tabular neighborhood, and a \textit{longitude} is a  1-cycle going once along the link component according to its framing. In other words, $a_I$ is the linking number between the longitude and the corresponding link component.}. The number of components of the link $\CL(\Gamma)$ is equal to the cardinality of $V$.

Equivalently, one can construct $M^3$ by associating to each vertex $I$ a copy of a lens space $L(a_I,1)\cong S^3/\Z_{a_I}$, understood as a circle fibration over $S^2$ with first Chern number equal to $a_I$.  If an edge connects two vertices, one punctures both fibrations at a point on the base and then glues both fibrations together by swapping the fiber circles with the circles surrounding the punctures on the bases.

In the following, we will restrict ourselves to the case when the graph is connected and there are no loops. 
Denote by $L$ the number of vertices of $\Gamma$ (i.e. $L=|V|$) and by $B$ the $L\times L$ linking matrix of $\CL(\Gamma)$ with entries
\begin{equation}
B_{IJ}=\left\{
\begin{array}{ll}
1,& I,J\text{ connected}, \\
a_I, & I=J, \\
0, & \text{otherwise}.
\end{array}
\right.\qquad I,J \in V.
\label{linking}
\end{equation}
In what follows we will denote by $\Pi$ the number of positive eigenvalues of $B$.  The matrix $B$ contains basic homotopy invariants of the 3-manifold. In particular, the first homology group of $M^3$ is given by the cokernel of the linking matrix, understood as a linear map $B:\Z^L\rightarrow \Z^L$
\begin{equation}
H_1(M^3,\Z) \cong \mathrm{Coker} B = \Z^L/B\Z^L.
\label{H1-coker}
\end{equation}
Assume for simplicity that $B$ is nondegenerate so that $\mathrm{Coker} B$ is a finite abelian group. Then $M^3$ is a rational homology sphere, i.e. $b_1(M^3)=0$, and it has a natural linking pairing on the first homology group\footnote{In general the pairing is only defined on the torsion subgroup.}:
\begin{equation}
    \begin{array}{rccl}
         \lk: & H_1(M^3,\Z)\otimes H_1(M^3,\Z) & \longrightarrow & \Q/\Z, \\
         & [\gamma_1]\otimes [\gamma_2] & \longmapsto & \frac{\#(\gamma_1 \cap \beta_2)}{n}\mod 1\qquad (n\gamma_2=\partial \beta_2).  
    \end{array}
    \label{linking-pairing-def}
\end{equation}
Using the isomorphism (\ref{H1-coker}) the pairing can be expressed in terms of the linking matrix
\begin{equation}
\lk(a,b) =(a,B^{-1}b)\;\mod \Z,\qquad a,b\in  \Z^L/M\Z^L.
\end{equation}

As it is well known, Dehn surgeries on different framed links can result in homeomorphic 3-manifolds. This happens if and only if the links can be related by a sequence of the so-called  three-dimensional Kirby \cite{kirby1978calculus}, or, equivalently Fenn-Rourke moves \cite{fenn1979kirby}. 

\begin{figure}[tbp]
\centering 
		\includegraphics[scale=1.2]{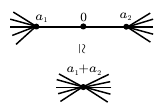}
		\includegraphics[scale=1.2]{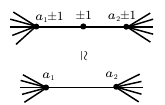}
		\includegraphics[scale=1.2]{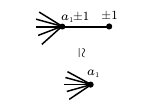}
\caption{Kirby-Neumann moves between plumbings that realize homeomorphic 3-manifolds.}
\label{fig:kirby-moves}
\end{figure}

For the framed links of the class considered above (see in particular Figure \ref{fig:plumbing}) the Kirby moves reduce to the so-called Neumann moves \cite{neumann1981calculus}. Specifically, two different plumbings $\Gamma$ and $\Gamma'$ realize two homeomorphic 3-manifolds if and only if they can be related by a sequence of moves depicted in  Figure \ref{fig:kirby-moves}. Therefore, if one defines a topological invariants of plumbed 3-manifolds in terms of the plumbing graph, it is sufficient to check its invariance under these basic Neumann moves.

\subsection{$\widehat{Z}$ of plumbed 3-manifolds for supergroups}
\label{sec:Zhat-plumbed-general}

Before introducing the homological block for plumbed $M^3$ and supergroup $\mathfrak{gl}(N|M)$, let us review the case $M=0$ considered in \cite{Gukov:2016njj,Gukov:2017kmk,Park:2019xey}. We will write the result in the following form\footnote{We will use the \textit{reduced} version of $\widehat{Z}$. The \textit{unreduced} version would have an extra universal $(q;q)_\infty^{-N}$ factor, where $(q;q)_\infty=\prod_{n=1}^{\infty}(1-q^n)$. On the level of underlying $Q$-cohomology, the removal of this factor corresponds to removal of a certain Fock space. \label{footnote-unreduced-glN}}, convenient for a generalization to the case of $M\neq 0$:
    \begin{multline}
	\widehat{Z}^{\mathfrak{gl}(N)}_a[M^3]= 
	(-1)^{\frac{N^2-N}{2}\Pi}
	q^{\frac{3\sigma-\text{Tr}M}{2}\frac{N^3-N}{12}}
	\int_\Omega 
	\prod_{I\in \text{Vert}}\prod_{\alpha=1}^N\frac{dz_{I\alpha}}{2\pi iz_{I\alpha}}
	\times
	\\ 
	\prod_{I\in \text{Vert}}
	\prod_{1\leq \alpha< \beta \leq N}\left(\sqrt{\frac{z_{I\alpha}}{z_{I\beta}}}-\sqrt{\frac{z_{I\beta}}{z_{I\alpha}}}\right)^{2-\deg(I)}
	\sum_{n\in (B\Z^{L})^N+\frac{a}{2}}
	q^{-\frac{1}{2}\sum_{\alpha=1}^N n^T_\alpha B^{-1}n_\alpha}\,
	\prod_{I,\alpha} z_{I\alpha}^{n_{I\alpha}}\,.
	\label{Zhat-plumbed-glN}
\end{multline}
Let us elaborate on the various elements of this formula. The indices $\alpha,\beta=1\ldots N$ and $I,J\in \vert$. The $\deg(I)\equiv \sum_{J\neq I}B_{IJ}$ denotes the degree of the vertex $I$, and $\sigma$ is the signature of the matrix $B$. 

The integration variables $z_{I\alpha}$ correspond to the eigenvalues of the holonomy of the $U(N)$ gauge field in the effective 3d theory on $D^2\times S^1_\text{time}$ associated to each vertex $I$. As described above, a single vertex in the plumbing by itself corresponds to a copy of the lens space $L(a_I,1)$, a circle fibration over $S^2$. By first reducing the fivebranes on the fiber one obtains a 5d $U(N)$ gauge theory, which one can further compactify on the $S^2$ base to obtain a 3d $U(N)$ gauge theory. The contour in $\Omega$ in (\ref{Zhat-plumbed-glN}) is chosen to be the principle value regularization of the contour $|z_{I\alpha}|=1$. That latter is the contour that naturally appears in the localization of 3d $\mathcal{N}=2$ theories on $D^2\times S^1$ \cite{Yoshida:2014ssa}. The principle value regularization in practice means that the formula should be understood as follows. The rational function in the variables $z_{I\alpha}$ in the first part of the integral should be expanded into power series in the chambers $|z_{I\alpha}/z_{I\beta}|<1$ or $|z_{I\alpha}/z_{I\beta}|>1$, according to where the contour of integration passes. For the principle value contour one has to take the average of the results corresponding to expansion in all possible chambers (related by $S_N$ Weyl symmetry action) for each vertex $I$. Note that the integrand itself is a well defined function in $q$ only when the matrix $B$ is negative definite. However, if one defines the integration by the procedure above, the result is still a well defined $q$-series if the plumbing graph satisfies a weaker condition of \textit{weak positivity} \cite{Gukov:2019mnk}. Namely, the condition that the matrix $B^{-1}$, restricted on the subspace generated by the vertices of degree greater than two, is negative definite. 

The index $a$, defined as the shift of the summation range, naturally belongs to $(\mathrm{Coker}\,2B)^N$, not $(\mathrm{Coker}\,B)^N\cong H_1(M^3,\Z)^N$. However $\widehat{Z}_a$ vanishes identically unless $a_{I\alpha}=(N-1)\,\deg(I)\mod 2,\;\forall \alpha$. When $N$ is even the corresponding subset of $\mathrm{Coker}\,2B$ is canonically isomorphic to $\text{Spin}^c(M^3)$ \cite{Gukov:2019mnk} (see also \cite{Gukov:2020frk,Gukov:2020lqm}), the set of spin$^c$ structures on $M^3$, which is non-canonically (unless one fixes a spin structure on $M^3$) isomorphic to $H_1(M^3,\Z)$. When $N$ is odd this subset of $\mathrm{Coker}\,2B$ is canonically isomorphic $\mathrm{Coker}\,B\cong H_1(M^3,\Z)$.

The sign and the overall power of $q$ in front of the integral (\ref{Zhat-plumbed-glN}) corresponds to the overall shift of the fermion parity and $L_0$ charge in the corresponding Hilbert space $\mathcal{H}_{a}$. It can be fixed for example by requiring invariance under the changes of plumbing that preserves the homeomorphism class of $M^3$. Note that in more invariant terms, applicable to the case of arbitrary Lie algebra, $(N^2-N)/2=|\Delta_+|$, where $\Delta_+$ is the set of positive roots and  $(N^3-N)/12=(\rho,\rho)$, where $\rho$ is the Weyl vector, that is half the sum of the positive roots. Finally let us note that the $\mathfrak{sl}(N)$ version can be obtained by restricting the integration to the subspace with $\prod_{\alpha=1}^N z_{I\alpha}=1,\; \forall I$.

Now, having reviewed the result for $U(N)$, we are ready to state its generalization for $U(N|M)$:
   \begin{multline}
	\widehat{Z}^{\mathfrak{gl}(N|M)}_{a,b}[M^3]= 
	(-1)^{\frac{(N+M)^2-(N+M)}{2}\Pi}
	q^{\frac{3\sigma-\text{Tr}M}{2}\frac{(N-M)^3-(N-M)}{12}} \times \\
	\int_\Omega 
	\prod_{I\in \text{Vert}}\prod_{\alpha=1}^N\frac{dz_{I\alpha}}{2\pi iz_{I\alpha}}
	\prod_{\alpha'=1}^M\frac{dy_{I\alpha'}}{2\pi iz_{I\alpha'}} \times
	\\ 
	\prod_{I\in \text{Vert}}
	\prod_{1\leq \alpha< \beta \leq N}\left(\sqrt{\frac{z_{I\alpha}}{z_{I\beta}}}-\sqrt{\frac{z_{I\beta}}{z_{I\alpha}}}\right)^{2-\deg(I)}
	\sum_{n\in (B\Z^{L})^N+\frac{a}{2}}
	q^{-\frac{1}{2}\sum_{\alpha=1}^N n^T_\alpha B^{-1}n_\alpha}\,
	\prod_{I,\alpha} z_{I\alpha}^{n_{I\alpha}}
	\times 
	\\
		\prod_{I\in \text{Vert}}
	\prod_{1\leq \alpha'< \beta' \leq M}\left(\sqrt{\frac{y_{I\alpha'}}{y_{I\beta'}}}-\sqrt{\frac{y_{I\beta'}}{y_{I\alpha'}}}\right)^{2-\deg(I)}
	\sum_{m\in (B\Z^{L})^M+\frac{b}{2}}
	q^{\frac{1}{2}\sum_{\alpha'=1}^M m^T_{\alpha'} B^{-1}m_{\alpha'}}\,
	\prod_{I,\alpha'} y_{I\alpha'}^{m_{I\alpha'}}
	\times 
	\\
	\prod_{I\in \text{Vert}}
	\prod_{\alpha=1}^N\prod_{\alpha'=1}^M\left(\sqrt{\frac{y_{I\alpha'}}{z_{I\alpha}}}-\sqrt{\frac{z_{I\alpha}}{y_{I\alpha'}}}\right)^{\deg(I)-2}.
	\label{Zhat-plumbed-glNM}
\end{multline}
The lines 2--4 of this long formula are simply two copies of the integrand in (\ref{Zhat-plumbed-glN}), where in the second copy the replacements $N\rightarrow M$, $z_{I\alpha}\rightarrow y_{I\alpha'}$, $n\rightarrow m$, $a\rightarrow b$, $q\rightarrow q^{-1}$ were made\footnote{If one used instead the \textit{unreduced} version of (\ref{Zhat-plumbed-glN}), the formula above would have an extra $(q;q)_\infty^{-N}\cdot (q^{-1};q^{-1})_\infty^{-M}$ factor (cf. Footnote \ref{footnote-unreduced-glN}). By using the relation $(x;q^{-1})_\infty = (xq,q)_\infty^{-1}$ one can bring it to the form $(q;q)_\infty^{-N}\cdot (1;q)_\infty^{-N}=(q;q)_\infty^{N-M}\cdot (1-1)^M$, which give a vanishing result, unless one removes the universal vanishing factor $(1-1)^N$. On the level of the underlying $Q$-cohomology this corresponds to factoring out the Fock space with $N$ fermionic generators with $L_0=0$.\label{foot-unreduced-super}}. The last replacement corresponds to the fact that $U(1)_q$ symmetry rotates the two stacks of branes in opposite directions, as described in Section \ref{sec:brane-setup}. These are the contributions of the degrees of freedom coming from each individual stack of fivebranes. In particular, the factors of the form $\prod_{1\leq \alpha < \beta \leq N}\ldots $ and $\prod_{1\leq \alpha' < \beta' \leq M}\ldots $ correspond to M2-branes stretching between pairs of different branes in the respective stacks. The choice of the contour $\Omega$ will be discussed later. 
 
The last line of (\ref{Zhat-plumbed-glNM}) represents the contribution of the degrees of freedom from intersections of two stacks of branes. It is simply a product of $N\cdot M$ copies of a single intersection. The latter was already considered in \cite{Gukov:2016gkn} in the case of $\mathfrak{gl}(1|1)$. 
As it is reflected in the dependence on the integration variables $z_{I\alpha}$ and $y_{I\alpha'}$, the contribution from the intersections of stacks of branes transform in the bifundamental representation of $U(N)\times U(M)$ gauge group associated to each vertex.

To illustrate this more concretely, consider the case when $\Gamma$ consists of a single vertex labelled by $p\in \Z$. Then the whole $M^3\cong L(p,1)$ is a circle fibration over $S^2_\text{base}$. The stacks of $N$ and $M$ fivebranes compactified on $M^3$ then can be effectively described by respectively $U(N)$ and $U(M)$ 3d $\CN=2$ gauge theories with adjoint chirals of R-charge 2 and Chern-Simons coupling $p$ \cite{Gadde:2013sca,Gukov:2016gkn} (note that the change of orientation on $D^2$, corresponding to the exchange $q\leftrightarrow q^{-1}$, is equivalent to the change of sign of $p$). As described earlier, these 3d theories can be obtained by reducing first the stacks of fivebranes on the circle fiber to get stacks of $N$ and $M$ D4-branes in Type IIA string theory and then by compactifying further the corresponding 5d $U(N)$ and $U(M)$ gauge theories on $S^2_\text{base}$, which has $p$ units of the Ramond-Ramond flux. These stacks of D4-branes intersect transversely along the $S^2_\text{base}\times S^1_\text{time}$. The theory living on the intersection is known to be a 3d hypermultiplet transforming in the bifundamental representation with respect to the gauge fields on the stacks of D4 branes. Its $S^2_\text{base}\times S^1_\text{time}$ (topologically twisted) index is exactly the contribution of a single vertex $I$ with $\deg(I)=0$ in the last line of (\ref{Zhat-plumbed-glNM}).

The overall constant and $q$-power in front of the integral in (\ref{Zhat-plumbed-glNM}) are the supergroup generalization of the corresponding factor in (\ref{Zhat-plumbed-glN}). In particular, $((N+M)^2-(N+M))/2=|\Delta_+|$ is the number of positive roots of $\mathfrak{gl}(N|M)$ (described explicitly below), which is the same as the total number of factors in the products in the integrand expression for a fixed $I$. The number $(N-M)^3-(N-M)=(\rho,\rho)$ is the square of the Weyl vector, i.e. $\rho = \rho_0 -\rho_1 =\frac{1}{2}\sum_{\alpha \in \Delta^+_{{0}}}\alpha -\frac{1}{2}\sum_{\alpha \in\Delta^+_{{1}}}\alpha$, the half-sum of the positive even roots minus the positive odd roots.

The indices $a$ and $b$ in (\ref{Zhat-plumbed-glNM}) are a priori valued in $\mathrm{Coker}(2B)^N$ and $\mathrm{Coker}(2B)^M$ respectively. However, similarly to the case of $\mathfrak{gl}(N)$ Lie algebra, $\widehat{Z}_{a,b}$ vanish identically unless $a_{\alpha I},b_{I\alpha'} = (N+M-1)\deg(I)\mod 2,\;\forall{\alpha,\alpha'}$. Therefore, when $(N+M)$ is odd, the indices effectively live in $H_1(M^3,\Z)^N$ and $H_1(M^3,\Z)^M$ respectively (modulo $S_N$ and $S_M$ permutations), and when $(N+M)$ is odd, they live in $\mathrm{Spin}^c(M^3)^N$ and $\mathrm{Spin}^c(M^3)^M$.

The $\mathfrak{sl}(N|M)$ version of (\ref{Zhat-plumbed-glNM}) can be obtained by restricting the integration to the subspace $\prod_{\alpha=1}^Nz_{I\alpha}=\prod_{\alpha'=1}^My_{I\alpha'},\;\forall I$. The $\mathfrak{psl}(N|N)$ version can be obtained by further restricting to the subspace $\prod_{\alpha=1}^Nz_{I\alpha}=1,\;\forall I$.

The expression (\ref{Zhat-plumbed-glNM}) and its $\mathfrak{sl}(N|M)$ and $\mathfrak{psl}(N|N)$ versions, can also be recast into the following compact form in terms of standard data of the super Lie algebra $\mathfrak{g}=\mathfrak{gl}(N|M)$:
  \begin{multline}
	\widehat{Z}^{\mathfrak{g}}_{a}[M^3]= 
	(-1)^{|\Delta_+|\Pi}
	q^{\frac{3\sigma-\text{Tr}M}{2}(\rho,\rho)} 
	\int_\Omega 
	\prod_{I\in \text{Vert}} dh_I  	\times\\
	\times \prod_{I\in \text{Vert}} {\mathcal{D}}_{\mathfrak{g}}(\alpha, h_I)^{2-\deg(I)}
	\sum_{\mathbf{n}\in (B\Z^L \otimes \Lambda)+\frac{a}{2}}
	q^{-\frac{1}{2}\mathbf{n}^T(B^{-1}\otimes K)\mathbf{n}}\,
	 e^{\mathbf{n}(\oplus_I\,h_I)}
	\label{Zhat-plumbed-superg}
\end{multline}
where $h_I$ belongs to a maximal torus of the simply connected Lie supergroup corresponding to $\mathfrak{g}$, $dh_I$ is the normalized measure on it, $\Delta_+$ is the set of positive roots, $\Pi$ is the number of positive eigenvalues of the linking matrix $B$, $\Lambda$ is the root lattice, $K:\Lambda\otimes \Lambda \rightarrow \Z$ is the Killing form on it and 
\begin{equation}
{\mathcal{D}}_{\mathfrak{g}}(\alpha, h_I):=	\prod_{\alpha\in \Delta_+}\left(e^{\alpha(h_I)/2}-e^{-\alpha(h_I)/2}\right)^{\epsilon(\alpha)}
\end{equation}
is the super Weyl denominator (which appears in the Weyl formula for super characters) where $\epsilon(\alpha)=\pm 1$ for even/odd roots $\alpha$.
In this form, equation \eqref{Zhat-plumbed-superg} can be understood as the formal generalization of the formula for a Lie algebra described in \cite{Park:2019xey} to the case of a Lie superalgebra. 
 
Indeed, the expression \eqref{Zhat-plumbed-glNM} can be recovered from \eqref{Zhat-plumbed-superg} through the knowledge of the (distinguished) root system of $\mathfrak{gl}(N|M)$, which can be realized as follows. Consider the vector space $\R^{M+N}$ equipped with the standard indefinite scalar product $(\cdot , \cdot)$ of signature $(N,M)$. It has the standard basis $e_{i},\,i=1\ldots N$, $f_i,\,i=1\ldots M$ satisfying
 \begin{equation}
    \begin{array}{rcl}
        (e_i,e_j) & = & \delta_{ij}, \\
         (f_i,f_j) & = & -\delta_{ij}, \\
         (e_i,f_j) & = & 0. \\
    \end{array}
    \label{glMN-basis}
 \end{equation}
 The even positive roots are then $e_i-e_j,\;i<j$ and $f_i-f_j,\;i<j$. The odd positive roots are $e_i-f_j$. The Killing form is induced by the scalar product $(\cdot,\cdot)$ on $\R^{M+N}$.
  
 Note that the integrand in  (\ref{Zhat-plumbed-superg}) by itself is never a well defined function of $q$ and $h_I$ when the Killing form $K$ is indefinite (which is generically the case for super Lie algebras) since the sum over $\mathbf{n}$ is divergent. This poses a significant complication compared to the case of ordinary Lie algebras that was considered previously in the literature. However if one instead treats the integral expression formally and applies the calculation procedure described earlier (first expanding the integrand into formal power series in $e^{\alpha(h_I)}$ and then taking the constant term), it can still give as a result a  well defined series in $q$ for an appropriate choice of the integration contour $\Omega$ (equivalently, the choice of the expansion chamber). In this paper we will not make a general analysis on when this is possible. Instead, we will show that it is possible in the case $\mathfrak{g}=\mathfrak{sl}(2|1)$ for a certain class of plumbed manifolds.

\section{The case of $\mathfrak{sl}(2|1)$}
\label{sec:Zhat-sl21}

In this section we consider the case of $\mathfrak{g}=\mathfrak{sl}(2|1)$ in detail.
In the notations of (\ref{glMN-basis}) there are 3 positive roots:
\begin{equation}
    \begin{array}{c}
         e_1-e_2,  \\
         e_1-f_1, \\
         e_2-f_1.
    \end{array}
\end{equation}
The $\mathfrak{sl}(2|1)$ root lattice is a rank 2 indefinite lattice generated by those vectors.  Let us introduce the following coordinates\footnote{Not to be confused with $z$'s and $y$'s in (\ref{Zhat-plumbed-glNM}). In the notations of (\ref{Zhat-plumbed-glNM})  $y_I=z_{I1},\;z_I=z_{I2}$ for $N=2$ and $M=1$ and thus $\mathfrak{g}=\mathfrak{sl}(2|1)$.} on the maximal torus: $y_{I}:=e^{(e_1-f_1)(h_I)}$ and $z_I:=e^{(e_2-f_1)(h_I)}$. In the sum over $\mathbf{n}$ in (\ref{Zhat-plumbed-superg}) we correspondingly decompose $\mathbf{n}_I=n_I(e_1-f_1)+m_I(e_2-f_1)$, so that $(\mathbf{n}_I,\mathbf{n}_I)=-2n_Im_I$. The formula (\ref{Zhat-plumbed-superg}) then takes the form
\begin{multline}
	\widehat{Z}_{a,b}^{\mathfrak{sl}(2|1)}[M^3]=(-1)^{\Pi}\int_\Omega 
	\prod_{I\in \vert}\frac{dz_I}{2\pi iz_I}\left.\frac{dy_I}{2\pi iy_I}\,
	\left(\frac{y_I-z_I}{(1-z_I)(1-y_I)}\right)^{2-\deg(I)}
	\right.
	\times \\
	\sum_{\substack{n\in B\Z^L+a \\ m\in B\Z^L+b}}
	q^{\sum_{I,J}B^{-1}_{IJ}\,n_Im_J}\,
	\prod_{J} z_J^{m_J}y_J^{n_J}
	\label{Zhat-plumbed-sl21}
\end{multline}
where
\begin{equation}
    a,b\in \mathrm{Coker}\,B\cong H_1(M^3,\Z).
\end{equation}
In order for the result to be a well defined $q$-series, we have to carefully choose the contour $\Omega$. We remind that the choice of contour should be understood as the choice of \textit{chamber} in the space of integration variables where the rational function 
\begin{equation}
    \prod_{I\in \vert}
	\left(\frac{y_I-z_I}{(1-z_I)(1-y_I)}\right)^{2-\deg(I)}
	\label{sl21-vert-prod}
\end{equation}
inside the integral is expanded. One should also consider the possibility of taking a linear combination of the contributions from different chambers (as in particular it happens in the case of ordinary Lie algebras). 

For each factor in (\ref{sl21-vert-prod}) corresponding to a high-valency vertex $I$ ($\deg(I)>2$) there are two chambers in the space of $(y_I,z_I)\in (\C^*)^2$ with different power series expansions: $|z_I|>|y_I|$ or $|z_I|<|y_I|$. For each factor associated to a vertex $I$ with $\deg(I)<2$ there are four chambers corresponding to all possible combinations of inequalities $|y_I|>1$, $|y_I|<1$ and $|z_I|>1$, $|z_I|<1$. The vertices with $\deg(I)=2$ contribute a trivial factor to the product above and thus there is a unique chamber for the corresponding variables $(y_I,z_I)$. Together these combine to give all possible chambers in the space of all integration variables $\{(y_I,z_I)\}_{I\in \vert} =(\C^*)^{2L}$. Therefore, a priori there are multiple possibilities of inequivalent choices of contour $\Omega$, growing with the complexity of the plumbing graph. 

We will call a chamber \textit{good} if its contribution results in a well defined $q$-series. Namely, this means the following. In a fixed chamber consider expansion up to a fixed order $d$ in each  variable $z_I^{\pm 1},y_I^{\pm 1},\;I\in \vert$. This truncated expansion is a Laurent polynomial. Multiply by it the infinite sum in the second line of (\ref{Zhat-plumbed-sl21}), which should be considered as the formal power series in $y_I$ and $z_I$. Then take the constant term of this new formal power series and multiply it by $(-1)^\Pi$. The result is a polynomial in $q$, up to an overall rational power of $q$, that is an element of $q^{\Delta_{ab}}\Z[q]$, for some $\Delta_{ab}\in \Q$. The chosen chamber is then called \textit{good} if this polynomial, with the constant term removed (if present),  stabilizes\footnote{Meaning that for an arbitrary large positive integer $K$, there exists a large enough expansion order $d$ so that the corresponding element of $q^{\Delta_{ab}}\Z[[q]]$ and the polynomial coincides modulo $q^{K+\Delta_{ab}}\Z[[q]]$.} to an element of $q^{\Delta_{ab}}\Z[[q]]$ as $d\rightarrow \infty$. Here by $q^{\Delta_{ab}}\Z[[q]]$ we mean the space of (a priori formal) power series in $q$ with integer coefficients, up to an overall rational power of $q$. The index $\widehat{Z}_{a,b}$ then can be defined, up to a constant term, as a certain linear combination of the resulting elements in $q^{\Delta_{ab}}\Z[[q]]$ over the good chambers. The value of the rational shift $\Delta_{ab}$ depends non-trivially on the indices $a,b\in \mathrm{Coker}\,B \cong H_1(M^3,\Z)$. In particular,
\begin{equation}
    \Delta_{ab}\mod 1 \,= \, a^TB^{-1}b\mod 1 \, =\,  \lk(a,b).
\end{equation}
The constant terms in $\widehat{Z}_{a,b}^{\mathfrak{sl}(2|1)}$  which were ignored in the analysis above, in general do not stabilize. Since they require a special care, we will come back to this point at the end of the section.  Note that a constant term can only be present when $\Delta_{ab} \equiv 0\mod 1$.

In Appendix \ref{app:sl21-contour} we examine the existence and the choice of good chambers for a \textit{generic} plumbing (see below for the definition); we argue that for a generic plumbing, if a good chamber exists, there are only two good chambers that are related by an obvious symmetry $z_I\leftrightarrow y_I$ (simultaneously for all $I\in \vert$) and thus produce identical series. In particular, the exchange $z_I\leftrightarrow y_I$ does not give rise to an extra sign because  $\sum_{I\in \vert} \deg(I)$ is even, as it is equal to twice the number of edges. 
 Below we summarize the results of the analysis reported in Appendix \ref{app:sl21-contour}. 

Denote by $\vert|_{\deg=d}$ the subset of vertices with degree equal to $d$, where the equality can also be replaced by an inequality. Assume that $\vert|_{\deg>2}\neq \emptyset$, that is there is at least one vertex of degree greater than two. Then a good chamber, as defined above, exists if 
there exists a vector
\begin{equation}
    \alpha_I =\pm 1, \quad I\in \vert|_{\deg\neq 2}
\end{equation}
such that
\begin{equation}
     X\text{ is \textit{copositive},}\quad X_{IJ}:=-B^{-1}_{IJ}\alpha_I\alpha_J,\quad I,J\in \vert|_{\deg>2},
    \label{plumbing-cond-1}
\end{equation}
and
\begin{align}
    &\alpha_{I}\alpha_J B^{-1}_{IJ}\leq 0,  \qquad \forall \; I\in \vert|_{\deg=1},\qquad J\in\vert|_{\deg\neq 2}, \label{plumbing-cond-2} \\
    &\alpha_{I}\alpha_J B^{-1}_{IJ}< 0, \qquad  \forall \; I,J\in \vert|_{\deg=1},\quad I\neq J. \label{plumbing-cond-3}
\end{align}
By definition, the matrix $X$ is called \textit{copositive} if for any vector $v$ such that $v_I\geq 0,\,\forall I$, with at least one $v_I\neq 0$, we have $\sum_{I,J}v_Iv_IX_{IJ}> 0$. A necessary and sufficient condition for this is that any principal submatrix of $X$ does not have a negative eigenvalue with the corresponding eigenvector having all positive components \cite{kaplan2000test}.

For a given vector $\alpha \in \{\pm 1\}^{V|_{\deg\neq 2}}$ that satisfies the above conditions there is a corresponding good chamber specified by the following inequalities:
\begin{equation}
\text{chamber $\alpha$:}
\qquad\qquad
    \begin{array}{cc}
        \deg(I)=1: &
        \left\{\begin{array}{c}
            |y_I|^{\alpha_I}<1, \\
            |z_I|^{\alpha_I}>1, \\
        \end{array}
        \right.
        \\
        \\
        \deg(I)>2: &
            \left|\frac{y_I}{z_I}\right|^{\alpha_I}<1.
    \end{array}
    \label{chamber-alpha}
\end{equation}
We define a plumbing to be \textit{generic} if there is at least one vertex with degree greater than two and there does not exist a non-trivial splitting $\vert|_{\deg\neq 2}=U\sqcup W$ into disjoint subsets $U$ and $V$, such that $B^{-1}_{IJ}=0$ if $I\in U$ and $J\in V$. It is then easy to see that if there exists $\alpha$ satisfying the conditions (\ref{plumbing-cond-1})-(\ref{plumbing-cond-3}), the inequalities (\ref{plumbing-cond-2})-(\ref{plumbing-cond-3}) alone fix it uniquely up to simultaneous change of signs $\alpha_I \leftrightarrow -\alpha_I$ for all $I$. This twofold ambiguity is due to the obvious exchange symmetry $y_I\leftrightarrow z_I$ in the integrand of (\ref{Zhat-plumbed-sl21}), originating from the $\Z_2$ Weyl symmetry of $\mathfrak{sl}(2|1)$. The contribution from these two chambers are then identically equal. 

For a generic plumbing admitting $\alpha$ that satisfies the conditions  (\ref{plumbing-cond-1})-(\ref{plumbing-cond-3}), one can finally write the following  formula which defines the $\mathfrak{sl}(2|1)$ homological blocks unambiguously (up to a constant term, which will be fixed below), 
\begin{multline}
	\widehat{Z}_{a,b}^{\mathfrak{sl}(2|1)}=(-1)^{\Pi}
	\mathrm{CT}_{y,z}\left\{
	\left.
	\left(\frac{y_I-z_I}{(1-z_I)(1-y_I)}\right)^{2-\deg(I)}
	\right|_{\text{chamber }\alpha}
	\right.
	\times\\
	\left.
	\sum_{\substack{n\in B\Z^L+a \\ m\in B\Z^L+b}}
	q^{\sum_{I,J}B^{-1}_{IJ}\,n_Im_J}\,
	\prod_{J} z_J^{m_J}y_J^{n_J} 
	\right\} \\ \;\; \in  q^{\Delta_{ab}}\Z[[q]]
	\label{Zhat-plumbed-sl21-CT}
\end{multline}
where $\text{CT}_{z,y}$ denotes the operation of taking the constant term of the formal power series in $y_I$ and $z_I$. It is easy to see that the resulting $q$-series are not just formal, but convergent in $|q|<1$ domain, as the coefficients grow at most polynomialy. 

In the following, we deal with the possible constant terms generated by the expression (\ref{Zhat-plumbed-sl21-CT}). As was previously pointed out (and explained in more detail in Appendix \ref{app:sl21-contour}), the term $\propto q^0$ produced by the formula (\ref{Zhat-plumbed-sl21-CT}) is in general ill-defined, as it is an infinite sum of integers. We will redefine it using the standard $\zeta$-regularization procedure. As we will see, this regularization do not spoil inveriance under Kirby-Neumann moves and it is also consistent with the conjectural relation to the invariants associated to a quantum supergroup (see Section \ref{sec:quantum-supergroup} and Appendix \ref{app:gauss-sums}).

In the first line of equation (\ref{Zhat-plumbed-sl21-CT}), in the product over vertices we substitute
\begin{equation}
    \begin{array}{c}
        y_I\longrightarrow y_I\,e^{-\alpha_I\epsilon}, \\
        z_I\longrightarrow z_I\,e^{\alpha_I\epsilon}.
    \end{array}
\end{equation}
In this way, the constant term produced by (\ref{Zhat-plumbed-sl21-CT}) can be represented as a series of the form
\begin{equation}
    \sum_{n\geq 0} c_n e^{-n\epsilon}\in \Z[[e^{-\epsilon}]].
    \label{epsilon-series}
\end{equation}
Moreover, this series can be always summed up to rational function in $e^{-\epsilon}$, which in general has singularity at $\epsilon=0$. We then define the constant term of $\widehat{Z}_{a,b}$ as the constant term (i.e. $\propto \epsilon^0$) in the expansion of this rational function with respect to small $\epsilon$. Equivalently, one can consider the Mellin transform of the series (\ref{epsilon-series}):
\begin{equation}
    \sum_{n\geq 0} \frac{c_n}{n^s}.
\end{equation}
This series is convergent for $\mathrm{Re}\,s>1$ and the result can be expressed as a linear combination of Hurwitz zeta functions. The constant term of $\widehat{Z}_{a,b}^{\mathfrak{sl}(2|1)}$ is then given by the value of this linear combination at $s=0$. 

The constant term defined in this way is a rational number, and thus, we have
\begin{equation}
   \widehat{Z}_{a,b}^{\mathfrak{sl}(2|1)}
   \; \in \;
   \Q +q^{\Delta_{ab}}\Z[[q]].
\end{equation}

Using the formula (\ref{Zhat-plumbed-sl21-CT}) one can explicitly check that the $q$-series defined by the formula (\ref{Zhat-plumbed-sl21-CT}) in terms of the plumbing data are invariant under Kirby-Neumann moves (reviewed in Section \ref{sec:plumbed-3-manifolds}) and therefore define a topological invariant of plumbed 3-manifolds, as predicted by physics. Here we do not give a proof of this claim, however, it should essentially follow from the arguments given in \cite{Gukov:2019mnk,Park:2019xey} in the case of ordinary Lie algebras.

In the rest of the section we provide several examples of the explicit use of the formula (\ref{Zhat-plumbed-sl21-CT}), starting with the most basic 3-manifold, $S^3$.

\subsection{3-sphere}
\label{sec:ex-sphere}
\begin{figure}[tbp]
\centering 
\raisebox{-9ex}{\includegraphics[scale=1]{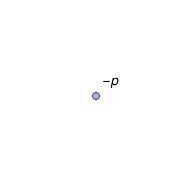}}
$\simeq$\raisebox{-14ex}{\includegraphics[scale=0.8]{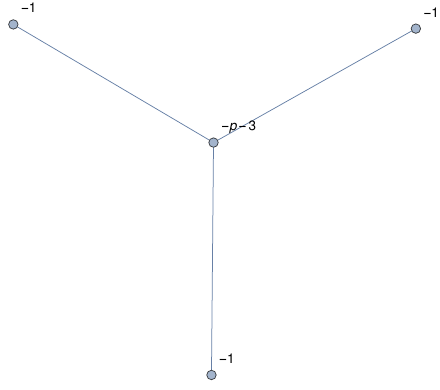}}
\caption{The left-hand side shows a common plumbing, containing a single vertex, that realizes a lens space $L(p,1)$. This plumbing however is non-generic, according to our definition. By a sequence of Kirby-Neumann moves, it can be transformed into the generic plumbing depicted on the right-hand side. }
\label{fig:lens-plumbing}
\end{figure}

It is possible to realize $S^3$, for example, via a  \textit{generic} plumbing shown on the right-hand side of Figure 
\ref{fig:lens-plumbing} (assuming $p=1$).
One can also use a non-generic, but much simpler plumbing, that consists of a single vertex labeled by $-1$. The latter is shown on the left-hand side of Figure 
\ref{fig:lens-plumbing} (for $p=1$).  The resulting $q$-series are the same for both realizations.

The generic plumbing has the following linking matrix and inverse linking matrix:
\begin{equation}
    B=\left(
        \begin{array}{cccc}
        -4 & 1 & 1 & 1 \\
        1 & -1 & 0 & 0 \\
        1 & 0 & -1 & 0 \\
        1 & 0 & 0 & -1 \\
        \end{array}
    \right),\qquad
    B^{-1}=\left(
        \begin{array}{cccc}
        -1 & -1 & -1 & -1 \\
        -1 & -2 & -1 & -1 \\
        -1 & -1 & -2 & -1 \\
        -1 & -1 & -1 & -2 \\
        \end{array}
    \right).
\end{equation}
The unique (up to an overall sign) solution to the constraints (\ref{plumbing-cond-1})-(\ref{plumbing-cond-3}) is provided by
\begin{equation}
    \alpha=\pm(1,1,1,1).
\end{equation}
As $H_1(S^3,\Z)=0$, there is a unique homological block. The application of the formula (\ref{Zhat-plumbed-sl21-CT}) then gives the following expression: 
	 \begin{multline}
		 \widehat{Z}^{\mathfrak{sl}(2|1)}[S^3]=1+2\zeta(0)+2\zeta(-1)+2\sum_{n\geq 1}\frac{q^n}{1-q^n}=
		 -\frac{1}{6}+2\sum_{m\ge 1}d(m)q^m
		 =\\
		 =-\frac{1}{6}+2(q+2q^2+2q^3+3q^4+2q^5+4q^6+2q^7+\ldots)
		 \label{Zhat-sphere}
	\end{multline}
	 	where $d(m)$ is the number of divisors of $m$. Note that in general the series of the form $\sum_{n}a_nq^n/(1-q^n)$ are known as Lambert series. The particular series in (\ref{Zhat-sphere}), up to an overall factor and a constant shift, are equal to the Eisenstein series of weight one,
	 	\begin{equation}
	 	\label{eq:eisentein1}
	 	    G_1(\tau):= \frac{1}{2} \zeta (0)+\sum_{m\ge 1}d(m)q^m\,.
	 	\end{equation}
	 	We review their quantum modular properties and resurgence properties in Sections \ref{sec:q-mod-sphere} and \ref{sec:resurgence-3sphere} respectively.

\subsection{Lens spaces}
The computation above for $S^3$ can be easily generalized to the case $M^3=L(p,1)$ ($p>0$), using either of the plumbings shown in Figure \ref{fig:lens-plumbing}. 
\label{sec:lens} 
In this case $H_1(L(p,1),\Z)\cong \Z_p$ and there are $p^2$ homological blocks labelled by pairs $(b,c)\in \Z_p$, then (\ref{Zhat-sphere}) generalizes to the following
	 \begin{equation}
		 \widehat{Z}^{\mathfrak{sl}(2|1)}_{b,c}[L(p,1)]=\mathrm{const}_{b,c}+2\sum_{m>0}d(m;p,b,c)q^{m/p}
		 \qquad \in q^{\frac{bc}{p}}\Z[[q]]
		 \label{Zhat-lens}
	\end{equation}
	 	where $d(m;p,b,c)$ is the number of positive integer pairs $(r,s)$ satisfying
		\begin{equation}
			\left\{\begin{array}{rcl}
				r &=& b \mod p, \\
				s &=&c \mod p, \\
				rs&=&m,
				\end{array}
			\right.
		\end{equation}
and $\mathrm{const}_{b,c}\in \Q$ denotes the constant term:
\begin{equation}
    \mathrm{const}_{b,c}=
    \left\{
        \begin{array}{rcl}
        1+2p\zeta(-1)+2\zeta(0), & &
        b=c=0\mod p, \\
 p\zeta(-1,b/p)+\zeta(0,b/p), & &
 c=0\mod p,\,b\neq 0\mod p,\\
p\zeta(-1,c/p)+\zeta(0,c/p), & &
 b=0\mod p,\,c\neq 0\mod p,\\
0, & &
b,c \neq 0 \mod p.
        \end{array}
    \right.
\end{equation}
where $\zeta(s,x)$ is the Hurwitz zeta function (see equation \eqref{eq:hurwitz_zeta}). The Weyl symmetry implies $\widehat{Z}^{\mathfrak{sl}(2|1)}_{b,c}=\widehat{Z}^{\mathfrak{sl}(2|1)}_{c,b}$, which is in agreement with the definition of $d(m;p,b,c)$ above. 

One can also rewrite (\ref{Zhat-lens}) more explicitly as follows,
 \begin{equation}
\widehat{Z}^{\mathfrak{sl}(2|1)}_{b,c}[L(p,1)] = \mathrm{const}_{b,c} + 2q^{\frac{(p-b)(p-c)}{p}-(p-b)}\sum_{k\ge 1} \frac{q^{ck}}{1-q^{pk-(p-b)}},
\label{Zhat-lens-explicit}
\end{equation}
taking $1\leq b,c\leq p$. Note that in the particular case $b,c=0\mod p$ the lens space homological block is simply related to the one of the 3-sphere:
\begin{equation}
    \widehat{Z}^{\mathfrak{sl}(2|1)}_{0,0}[L(p,1)]
    =\widehat{Z}^{\mathfrak{sl}(2|1)}[S^3]|_{q\rightarrow q^p}.
\end{equation}

A few remarks are in order. First, already when $M^3=L(p,1)$ the $\mathfrak{sl}(2|1)$ homological blocks are qualitatively quite different from the homological blocks associated to ordinary Lie algebras. In the latter case they are \textit{polynomials} in $q$ (up to an overall rational power), while in the former case they are full fledged $q$-series. Second, we note that the coefficients $d(m;p,b,c)$ in the $q$-series above coincide with the Euler characteristic of the moduli space of $m$ $SO(3)$ instantons on $L(p,1)\times \R$ propagating between flat connections labelled by $b\pm c$ \cite{austin1990so}. We expand further on this observation in Section \ref{sec:open-questions}. In Section (\ref{sec:q-mod-lens}) we will study quantum modularity properties of the $q$-series (\ref{Zhat-lens}). 

Finally, the {\textit{unreduced}} homological blocks of lens spaces can be related to characters of a particular sum of atypical modules $\widehat{\mathcal{A}}_{n,\ell}$, associated to the affine Kac-Moody superalgebra $\widehat{\mathfrak{gl}}(1|1)$ \cite{creutzig2011walgebras}. The homological blocks of lens spaces can be, in fact, related to 
\begin{equation}
\sum_{m \ge 1}\chi_{\widehat{\mathcal{A}}_{mn+1/2,m \ell}}(y,z;q)= z\sum_{m \ge 1} \frac{y^{m\ell} z^{mn}q^{(mn+1/2)m\ell +m^2\ell^2/2}}{1+zq^{m\ell}} \prod_{i=1}^{\infty}\frac{(1+zq^{i})(1+z^{-1}q^{i-1})}{(1-q^i)^2} 
\label{svoa-char}
\end{equation}
if we take $z=-q^{b-p}$, $y=(-1)^{1/2}q^{a/2}$ where $a= -1+b-p+2c/p$ and restrict to $n=-\ell/2$. 
This choice of $n$ singles out the modules whose affine highest weight states have conformal dimension zero.
Note that the infinite product in (\ref{svoa-char}) can be in principle attributed to the extra factor appearing in the \textit{unnormalized} version of the homological blocks (see Footnote \ref{foot-unreduced-super} for details).

\subsection{Seifert 3-manifolds with 3 exceptional fibers}
\label{sec:ex-seifert}
\begin{figure}[tbp]
\centering 
\includegraphics[scale=1]{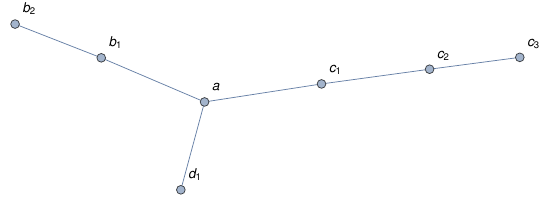}
\caption{An example of a plumbing realizing a Seifert fibration over $S^2$ with 3 exceptional fibers.}
\label{fig:seifert-3-fiber-plumbing}
\end{figure}

Below we consider a particular class of 3-manifolds which can be realized by plumbings with a single vertex of degree 3 and no other vertex of degree greater than two. An example of such a plumbing is illustrated in Figure \ref{fig:seifert-3-fiber-plumbing}. Such 3-manifolds can be equivalently realized as Seifert fibrations over $S^2$ with (at most) 3 exceptional fibers.

Using the explicit formula (\ref{Zhat-plumbed-sl21-CT}) it is possible to argue that in all such cases the homological blocks $\widehat{Z}^{\mathfrak{sl}(2|1)}_{b,c}$ are linear combinations of the $q$-series $F(q;\alpha,\beta;A,B)q^\gamma$, where
	\begin{equation}
		F(q;\alpha,\beta;A,B):=\sum_{m\geq 0} \frac{q^{\alpha m^2+\beta m}}{1-q^{Am+B}}.
	\end{equation}
Before we illustrate two explicit examples, note the close relation of $F(q;\alpha,\beta,\gamma;A,B)$ to ``half'' of an higher level Appell-Lerch sum. The latter being defined as, 
\begin{equation}
    A_{\ell}(u,v;\tau):= y^{\ell/2}\sum_{n\in \Z} \frac{(-1)^{\ell n}z^n q^{\ell n(n+1)/2}}{1-yq^n}
\end{equation}
where $y=e^{2\pi i u}$, $z=e^{2\pi i v}$ and as usual $q=e^{2\pi i \tau}$. As in the case of lens spaces, the homolgical blocks of Seifert 3-manifolds with 3 exceptional fibers appear to be closely related to  a certain sum of atypical modules $\widehat{A}_{n,\ell}$ introduced in the previous section. 
See \cite{creutzig2011walgebras, alfes2012mock} for further details on W-algebras extending $\widehat{\mathfrak{gl}}(1|1)$ and their connection to Appell-Lerch sums.\footnote{Higher level Appell-Lerch sums also appear in connection to characters of $\mathcal{N}= 2$ minimal models in \cite{Creutzig_2019}.}.

\subsubsection{$\overline{\Sigma(2,3,5)}$}
\label{sec:ex-235}
\begin{figure}[tbp]
\centering 
\raisebox{-15ex}{\includegraphics[scale=1]{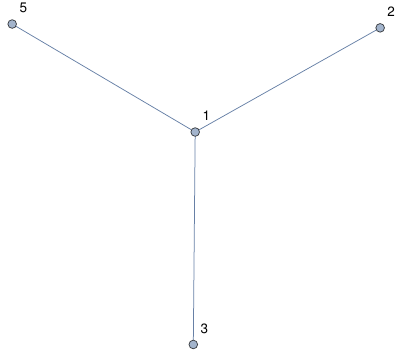}}$\simeq$ \raisebox{-6ex}{\includegraphics[scale=0.8]{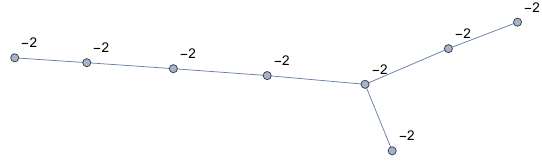}} 
\caption{Two equivalent plumbings (with respect to the Kirby-Neumann moves), which realize the Poincar\'e homology sphere, $\overline{\Sigma(2,3,5)}$.}
\label{fig:sigmabar235-plumbing}
\end{figure}  
Consider a 3-manifold realized by either of the plumbings shown in Figure \ref{fig:sigmabar235-plumbing}. This manifold is known as Poincar\'e homology sphere. It is also a particular example of a Brieskorn 3-sphere and thus can be denoted $\overline{\Sigma(2,3,5)}$. The bar indicated that the orientation is reversed compared to the standard one. 

The linking matrix of the plumbing on the left-hand side of Figure \ref{fig:sigmabar235-plumbing}, and its inverse, read
\begin{equation}
	B=\left(\begin{array}{cccc}
		1 & 1 & 1 & 1 \\
		1 & 2 & 0 & 0 \\
		1 & 0 & 3 & 0 \\
		1 & 0 & 0 & 5 
	\end{array}\right),
	\qquad
	B^{-1}=\left(\begin{array}{cccc}
		-30 & 15 & 10 & 6 \\
		15 & -7 & -5 & -3 \\
		10 & -5 & -3 & -2 \\
		6 & -3 & -2 & -1 
	\end{array}\right).
\end{equation}
The constraints (\ref{plumbing-cond-1})-(\ref{plumbing-cond-3}) are satisfied with $\alpha=\pm(1,-1,-1,-1)$.

As $H_1(\overline{\Sigma(2,3,5)},\Z)\cong 0$ there is a single homological block. Using the formula (\ref{vertex-series-degree-1}) and formula (\ref{vertex-series-high-degree}) for $K=1$
one can reduce the expression in (\ref{Zhat-plumbed-sl21-CT}) to a linear combination of quadruple semi-infinite sums. Moreover, in each term in this linear combination, the sum over two or three variables can be performed explicitly (using the formula for an infinite geometric sum). Therefore the final result can be written as a linear combination of single and double semi-infinite sums. After some manipulations (including shifting the summation variables) the result can be written as follows (up to a constant term): 
\begin{multline}
	\widehat{Z}^{\mathfrak{sl}(2|1)}[\overline{\Sigma(2,3,5)}]=  
	\sum_{m,n\geq 1} q^{15(m-1)(2m+n)}P_1(q^{2m+n})
	+\sum_{m,n\geq 1} q^{10(m-1)(3m+n)}P_2(q^{3m+n})
	\\
	+\sum_{m,n\geq 1} q^{6(m-1)(5m+n)}P_3(q^{5m+n})
	+
	\sum_{n\geq 1}\cfrac{P_4(q^n)}{1-q^n}
	+
	\sum_{m\geq 1}\cfrac{q^{30m(m-1)}\,P_5(q^m)}{1-q^m}
	\label{PHS-q-series}
\end{multline}
where $P_i$ are the following \textit{polynomials}:
\begin{equation}
\begin{array}{rl}
	P_1(x):=& -\cfrac{(1+x^8)(1-x^{15})^2}{(1-x^3)(1-x^5)}, \\
	P_2(x):=& -\cfrac{(1+x^7)(1-x^{10})^2}{(1-x^2)(1-x^5)}, \\
	P_3(x):=& -\cfrac{(1+x^5)(1-x^{6})^2}{(1-x^2)(1-x^3)}, \\	
	P_4(x):=& 3-3x+2x^2+2x^5-x^6+x^7+x^8-x^{10}\\
	&+2x^{11}-x^{12}+x^{13}-x^{15}+x^{16}, \\
	P_5(x):=& -\cfrac{(1-x)(1-x^{30})^2(1+x^{16}+x^{21}+x^{25}-2x^{31})}{(1-x^6)(1-x^{10})(1-x^{15})}.
\end{array}
\end{equation}
The first few terms in the $q$-expansion, including the constant term, are the following:
	\begin{equation}
		\widehat{Z}^{\mathfrak{sl}(2|1)}[\overline{\Sigma(2,3,5)}]
		= -\frac{1}{6}+2q^2+2q^3+4q^4+4q^5+6q^6+4q^7+\ldots
	\end{equation}
In Section \ref{sec:resurgence-235} we will analyze resurgence property of this series with respect to $\hbar:=-\log q$.

\subsubsection{$\Sigma(2,3,7)$}
\label{sec:ex-237}
\begin{figure}[tbp]
\centering 
\includegraphics[scale=1]{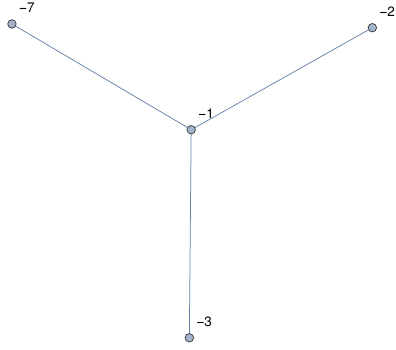}
\caption{A plumbing that realizes Brieskorn homology sphere ${\Sigma(2,3,7)}$.}
\label{fig:sigma237-plumbing}
\end{figure}

Consider the plumbing shown in Figure \ref{fig:sigma237-plumbing}. This manifold is known as Brieskorn 3-sphere $\Sigma(2,3,7)$.
We have again $H_1(\Sigma(2,3,7),\Z)\cong 0$ and thus there is a unique homological block. Similarly to the case of the Poincar\'e homology sphere the $q$-series can be explicitly written in the following form:
\begin{multline}
    \widehat{Z}^{\mathfrak{sl}(2|1)}[\Sigma(2,3,7)]=
	\sum_{m,n\geq 1} q^{21(m-1)(2m+n)}P_1(q^{2m+n})
	+\sum_{m,n\geq 1} q^{14(m-1)(3m+n)}P_2(q^{3m+n})
	\\
	+\sum_{m,n\geq 1} q^{6(m-1)(7m+n)}P_3(q^{7m+n})
	+
	\sum_{n\geq 1}\cfrac{P_4(q^n)}{1-q^n}
	+
	\sum_{m\geq 1}\cfrac{q^{42m(m-1)}\,P_5(q^m)}{1-q^m}
	\label{237-q-series}
\end{multline}
where $P_i$ are the following \textit{polynomials}:
\begin{equation}
\begin{array}{rl}
	P_1(x):=& -\cfrac{(1+x^{10})(1-x^{21})^2}{(1-x^3)(1-x^7)}, \\
	P_2(x):=& -\cfrac{(1+x^9)(1-x^{14})^2}{(1-x^2)(1-x^7)}, \\
	P_3(x):=& -\cfrac{(1+x^5)(1-x^{6})^2}{(1-x^2)(1-x^3)}, \\	
	P_4(x):=& x^{22}-x^{21}+x^{19}-x^{18}+x^{16}+x^{15}-2 x^{14}+2 x^{13}\\
	&-x^{11}+2 x^9-x^8+x^7+x^6+2 x^2-3 x+3, \\
	P_5(x):=& -\cfrac{(1-x)(1-x^{42})^2(1+x^{20}+x^{27}+x^{35}-2x^{41})}{(1-x^6)(1-x^{14})(1-x^{21})}.
\end{array}
\end{equation}
The first few terms in the $q$-expansion are the following:
	\begin{equation}
		\widehat{Z}^{\mathfrak{sl}(2|1)}[\Sigma(2,3,7)]
		= -\frac{1}{6}+2q^2+2q^3+4q^4+2q^5+6q^6+4q^7+\ldots
	\end{equation}
In Section \ref{sec:resurgence-237} we will analyze resurgence property of this series with respect to $\hbar:=-\log q$.

\section{Resurgence}

\label{sec:resurgence}

As reviewed in Section \ref{sec:brane-setup} (mostly following \cite{Witten:2011zz,Mikhaylov:2014aoa}), the M-theory setup, in which the BPS indices $\widehat{Z}^{\mathfrak{g}}_a[M^3]$ are naturally defined, is related by a sequence of dualities to the setup in type IIB theory realizing 3d supergroup Chern-Simons theory with gauge supergroup\footnote{As a reminder, for concreteness we assume that $G$ is simply-connected.} $G$ with Lie algebra $\mathfrak{g}$ on $M^3$, analytically continued with respect to the level. From this construction, one expects in particular that the homological blocks, $\widehat{Z}^{\mathfrak{g}}_a[M^3]$, labelled by different $a$ are related by a linear transformation to the path integral in the supergroup Chern-Simons theory on $M^3$ over the Lefschetz thimble contours\footnote{That is, unions of steepest descent paths from connected components of the critical set of the action functional.} in the space of connections. Such relation was explicitly verified in the case of ordinary group $G=SU(2)$ for certain manifolds in \cite{Gukov:2016njj} using techniques of resurgence theory.

However, a path integral approach to supergroup Chern-Simons theory has not yet been well developed except in the $G=U(1|1)$ case (see in particular \cite{Rozansky:1992zt,Mikhaylov:2015nsa}). One of the issues, that has been pointed out in \cite{Mikhaylov:2014aoa}, is that the naturally defined path integral of the theory on a compact 3-manifold, without any Wilson lines insertions, is infinite. This is due to the fact that one has to divide by the volume of the gauge group, $\mathrm{Vol}(G)$, which is generically zero for supergroups\footnote{The exceptions are $OSp(1|2n)$ supergroups.}. However, this issue can in principle be circumvented by removing a point from the 3-manifold and interpreting the removed point as the point at infinity, so that the resulting 3-manifold asymptotically looks like $\R^3$ (cf. \cite{Witten:2011zz}). Then one only needs to consider the gauge transformations that become trivial at infinity and $1/\mathrm{Vol}(G)$ factor does not arise. Equivalently, one can consider a compact 3-manifold and pick a base point $x_0\in M^3$ and consider equivalence classes of connections only with respect to \textit{based} gauge transformations, i.e. corresponding to the maps $M^3\rightarrow G$ such that $x_0\mapsto \mathbf{1}\in G$.

Barring such normalization issues aside and without explicit knowledge of the linear transformation between $\widehat{Z}^{\mathfrak{g}}_a[M^3]$ and the integrals over the Lefschetz thimbles, one can still perform nontrivial checks of this relation through resurgence theory, similarly to how it was done in \cite{costin2011resurgence,Gukov:2016gkn} in the case of $G=SU(2)$. In particular, if one considers the asymptotic expansion of $\widehat{Z}^{\mathfrak{g}}_a[M^3]$ at $q\rightarrow 1$, one should be able to detect the contributions of the different critical points of the Chern-Simons actions, that is flat connections on $M^3$. The limit $q\rightarrow 1$ is equivalent to sending the Chern-Simons level $k$ (related to $q$ as $q=e^{\frac{2\pi i}{k}}$) to infinity. This is nothing but the weak coupling limit and hence a perturbative description is available.    

Before we proceed to consider particular examples, let us very briefly review the basics of resurgence theory and fix some conventions. For a more comprehensive review, adjusted to the setting of analytically continued Chern-Simons theory with ordinary gauge group \cite{Witten:2010cx,Kontsevich}, we suggest \cite{Gukov:2016njj}. In the context of supergroup Chern-Simons theory, we will need a mild generalization of that setup. 

Assume the path integral of the theory can be formally treated as a finite dimensional integral of the form 
\begin{equation}
    Z(\hbar)= \frac{1}{\hbar^{m/2+c}}\int_{\Gamma} d^mx\,f(x) \,e^{-\frac{S(x)}{\hbar}}
    \label{resurgence-model-integral}
\end{equation}
where $S(x)$ is a holomorphic function of \textit{complex} variables $x$,  $\Gamma$ is a mid-dimensional contour, and $c$ is a constant depending on the chosen normalization of the integral. We allow $S(x)$ to have a degenerate, but not vanishing, matrix of second derivatives at the critical points. The function $f(x)$ in (\ref{resurgence-model-integral}) is only assumed to be meromorphic in $x$, unlike the more standard scenario when it is assumed to be holomorphic. The reason for this is the following. In the path integral of the supergroup Chern-Simons theory in principle there are also Grassmann valued fields, corresponding to odd components of the connection 1-form. However, one can do a partial gauge fixing down to the even subgroup (cf. \cite{Kapustin:2009cd,Mikhaylov:2014aoa}). This will result in the corresponding superghost fields that has even parity (unlike the usual ghosts that have odd parity). Integrating them out will lead to the inverse determinant of the quadratic form that defines the action for superghosts and depends on the bosonic gauge fields. In the finite dimensional model for the path integral this can be taken into account by allowing a non-trivial meromorphic $f(x)$ inside the integral (\ref{resurgence-model-integral}). It has singularities corresponding to bosonic connections for which the superghosts have zero-modes\footnote{A similar and related phenomenon appears in supermatrix models \cite{AlvarezGaume:1991zc,Yost:1991ht}. In particular, a Hermitian supermatrix can be gauge-fixed to a purely bosonic diagonal matrix. The integral over the space of Hermitian supermatrices modulo superunitary transformations then reduces to the integral over the eigenvalues with an insertion of a rational function of the eigenvalues (in the case of ordinary matrices the rational function is actually a polynomial). 

Note that a direct relation between $\widehat{Z}^{\mathfrak{g}}_a[M^3]$ and  supermatrix models appears in the case of $M^3=L(p,1)$. The expressions (\ref{Zhat-plumbed-glNM}) and (\ref{Zhat-plumbed-superg}) are then essentially equivalent to the supermatrix version \cite{Drukker:2009hy,Marino:2009jd,Drukker:2010nc} of the matrix model that appears in the localization of Chern-Simons theory on $L(p,1)$ \cite{lawrence1999witten,Marino:2002fk,Halmagyi:2003ze,Beasley:2005vf}.}. The function $S(x)$ then plays the role of a finite-dimensional analogue of the Chern-Simons action for maximal even subgroup of the supergroup $G$. In particular, when $G=SU(N|M)$ this subgroup is $SU(N)\times SU(M)\times U(1)$, up to a quotient over a finite subgroup. The ``Planck constant'' $\hbar$ is related to the analytically continued Chern-Simons level as $\hbar=-\frac{2\pi i}{k}=-\log q$. Note that in order to consider analytic continuation away from $k\in \Z$ one has to sacrifice the invariance under large gauge transformations. Equivalently, in the path integral one integrates over the universal cover of the original integration space, which is the space of connections modulo all gauge transformations. 
 
Without loss of generality, assume that $S(x)$ in (\ref{resurgence-model-integral}) has a critical point at $x=0$ with critical value $S(x_0)=0$. In Chern-Simons theory it will correspond to a trivial flat connection. Consider $\Gamma$ to be the Lefschetz thimble associated to this critical point, that is the union of steepest descent paths with respect to $-\mathrm{Re}\,S(x)/\hbar$.  Note that if $f(x)$ is singular anywhere on $\Gamma$ (in particular at $x_0$) one has to regularize the contour by shifting it infinitesimally away from the singular locus of $f(x)$. In general there are multiple ways to do this and they will produce different results. In the context of supergroup Chern-Simons theory the multiple choices of avoiding singularities of $f(x)$ in principle can be tracked down to the choice of contour in the integral involving both bosonic gauge fields and superghosts\footnote{A possibly simplest finite dimensional model for this phenomenon is the integral 
\begin{equation}
    \int dxdydz e^{-\frac{1}{\hbar}(x^2/2-xyz)}\propto \int dx\, \frac{1}{x} \,e^{-\frac{x^2}{2\hbar}} 
\end{equation}
where both sides have two linearly independent contours.
}. 

The integral (\ref{resurgence-model-integral}) then has an asymptotic series of the form:
\begin{equation}
	Z(\hbar) \; \approx \; \text{(singular terms in $\hbar$)}\; + \; \sum_{\substack{L\in \frac{1}{2}\Z \\ L\geq 0}}c_L \Gamma(L+1) \hbar^{L}
	,\qquad \hbar \rightarrow \infty.
	\label{Z-general-asympt}
\end{equation}
Define the Borel transform of the non-singular part of such series by
\begin{equation}
	B(\xi): = \sum_{\substack{L\in \frac{1}{2}\Z \\ L\geq 0}}c_L\xi^L.
	\label{B-general-series}
\end{equation}
This series has finite radius of convergence and can be analytically continued to at least a cover of $\C\setminus \{S_i\}_i$, a complex plane with a discrete set of points removed.  In other words, the analytically continued function $B(\xi)$ can have singularities (including possible branch points) only at isolated points $\xi=S_i$. The singularities can appear at critical values of $S(x)$. In the more general setting when $f(x)$ in (\ref{resurgence-model-integral}) is allowed to be meromorphic they can also appear at the critical values of $S(x)$ restricted to the singular loci of $f(x)$. Namely, one has to consider critical values of $S(x)$ restricted to codimension-1 singular loci of $f(x)$, codimension-2 singular loci (intersections of codimension-1 loci), codimension-3 singular loci and so on. In simple scenarios however the critical points of $S(x)$ restricted to singular loci may coincide with the critical points of the unrestricted $S(x)$ and therefore do not lead to any new positions $S_i$ for the singularities of $B(\xi)$. In this work we do not provide a detailed analysis of whether  such extra singularities in $B(\xi)$ can actually arise in the context of supergroup-Chern-Simons (this would require a careful analysis of bosonic Chern-Simons action restricted on the loci where superghosts have zero modes).  In the specific examples that we consider below, we only detect the singularities of $B(\xi)$ at the critical values of the unrestricted Chern-Simons functional.

The position of singularities of a locally holomorphic function $B(\xi)$ can be deduced from the asymptotics of its coefficients $c_L$ in (\ref{B-general-series}) at $L\rightarrow \infty$. In particular, if
\begin{equation}
	c_L = 
	\left\{\begin{array}{rl}
	\sum_i \mathcal{A}_i\,\frac{L^{\alpha_i}}{S_i^L}(1+O(L^{-1})),& L\in \Z, \qquad L\rightarrow
	\infty, \\
		0, & L\in \Z+\frac{1}{2},\\
	\end{array}\right.
\end{equation}
for some $\alpha_i \in \frac{1}{2}\mathbb{Z}$ and some non-vanishing complex constants $\mathcal{A}_i$ and $S_i$, then $B(\xi)$ has singularities at $\xi = S_i$ where it behaves as 
\begin{equation}
	B(\xi) = \frac{\mathcal{A}_i\,\Gamma(1+\alpha_i)S_i^{1+\alpha_i}}{(S_i-\xi)^{1+\alpha_i}}\,(1+O(\xi-S_i)),\qquad \xi\rightarrow S_i.
	\label{B-general-sing}
\end{equation}
If, on the other hand,
\begin{equation}
	c_L = 
	\left\{\begin{array}{rl}
	0, & L\in \Z,\\
	\sum_i \mathcal{A}_i\,\frac{L^{\alpha_i}}{S_i^L}(1+O(L^{-1})),& L\in \Z+\frac{1}{2}, \qquad L\rightarrow
	\infty
	\end{array}\right.
\end{equation}
then $B(\xi)$ has a degree 2 branch point at the origin and has singularities at $\xi = S_i$ on both branches where its leading behavior is still given by the formula (\ref{B-general-sing}), but with an overall $\pm 1$ sign depending on the choice of the branch. The more general case with $c_{L}\neq 0$ for both $L\in \Z$ and $L\in \Z+1/2$ can always be recovered from a linear combination of the two special cases above.

Moreover, the integral (\ref{resurgence-model-integral}) can be expressed exactly as\footnote{This choice of normalization will prove to be convenient later on.}
\begin{equation}
	Z(\hbar) \; = \; \text{(singular terms in $\hbar$)}\; + \; 
	\frac{1}{\hbar}\int_{\gamma}d\xi\,B(\xi) \,e^{-\xi/\hbar}
\end{equation}
where $\gamma$ is the one-dimensional contour corresponding to the multi-dimensional contour $\Gamma$ in (\ref{resurgence-model-integral}) via the projection $S: \C^{M} \rightarrow \C$, where $\C^M$ is the complex space parametrized by $x$. Suppose one starts deforming the  contour $\gamma$ in a continuous way. If at some point it crosses a singular point $\xi=S_i$, the result of the integral changes by $\Delta_i Z$, whose asymptotic expansion in $\hbar$ is given by
\begin{equation}
	\Delta_i Z = \frac{2\pi i \mathcal{A}_iS^{1+\alpha_i}_i\,e^{-S_i/\hbar}}{\hbar^{1+\alpha_i}}\,(1+O(\hbar^{1/2})).
	\label{stokes-jumps}
\end{equation}
Such changes are often referred to as \textit{Stokes jumps}.

The finite dimensional model then gives certain concrete predictions to the asymptotic behavior of $\widehat{Z}_{a,b}^{\mathfrak{sl}(2|1)}[M^3]$ at 
\begin{equation}
    q\equiv e^{-\hbar} \rightarrow 1-,
\end{equation}
assuming it is indeed related to some linear combination of contour integrals in the analytically continued supergroup Chern-Simons theory. The first prediction is that the asymptotic expansion should be such that its Borel transform $B(\xi)$ has finite radius of convergence and can be analytically continued beyond that. The second prediction (assuming the first one holds) is that the values of the Chern-Simons functional on flat connection of the complexified maximal even subgroup should appear as positions of the singularities of $B(\xi)$. In the case of $G=SU(2|1)$ the maximal even subgorup is\footnote{This can be seen by representing a general even superunitary matrix of type $2|1$ in the form
\begin{equation}
    \left(
        \begin{array}{c|c}
            V & \begin{array}{c} 0 \\ 0 \end{array} \\
            \hline
            \begin{array}{cc} 0 & 0 \end{array} & \det V
        \end{array}
    \right)
\end{equation} 
where $V\in U(2)$.
} $SU(2)\times_{\Z_2} U(1)\cong U(2)$. However the Killing form induced from the one on $\mathfrak{sl}(2|1)$ is not the standard Killing form on $\mathfrak{gl}(2)\cong \frak{sl}(2)\oplus \frak{gl}(1)$. Rather, the contribution from the  $\mathfrak{gl}(1)$ central subalgebra comes with a negative sign. The complexification of $U(2)$ is $GL(2,\C)$.  As usual, the flat connections correspond to the homomorphisms $\rho:\pi_1(M^3)\rightarrow GL(2,\C)$, up to conjugation. The corresponding singularities then can be encountered at $\xi=S_i$ where $S_i/(-4\pi^2)\mod 1$ is the value of the Chern-Simons invariant of $\rho$, normalized modulo 1. In general we expect to have different singularities corresponding to the same flat connection. This is due to the fact that in analytically continued Chern-Simons theory one integrates over a mid-dimensional contour in the universal cover of the ordinary space of connections modulo gauge transformations. 

In the rest of this section we check this predictions for certain specific 3-manifolds. For technical simplicity we consider only the cases of homology 3-spheres, that is when $H^1(M^3,\Z)=0$, so that there is a single homological block $\widehat{Z}_{a,b}^{\mathfrak{sl}(2|1)}[M^3]$ with $a=b=0$. Moreover, under this assumption there are no non-trivial homomorphisms $\pi_1(M^3)\rightarrow \C^*$ and the image of any $\rho:\pi_1(M^3)\rightarrow GL(2,\C)$ is contained inside the $SL(2,\C)$ subgroup. This follows from the fact that the composition of $\rho$ with the determinant homomorphism $\det:GL(2,\C)\rightarrow \C^*$ must be trivial. 

\subsection{3-sphere}
\label{sec:resurgence-3sphere}
As was shown in Section \ref{sec:ex-sphere} the $\mathfrak{sl}(2|1)$ homological block of 3-sphere is given by the following Lambert series:
\begin{equation}
	\widehat{Z}^{\mathfrak{sl}(2|1)}=-\frac{1}{6}+2\sum_{n=1}^\infty \frac{q^n}{1-q^n}=-\frac{1}{6}+2\sum_{n=1}^\infty d(n)q^n
\end{equation}
where $d(n)$ is the number of divisors of $n$. Up to an overall factor and a constant term it is equal to the Eisenstein series of weight one. It has the following asymptotic expansion as $\hbar\equiv \log q \rightarrow 0+$ \cite{bettin2013reciprocity,Bettin_2013,Banerjee_2017} (cf. also \cite{Dorigoni:2020oon} for a recent study of resurgence properties of Lambert series in a different context):
\begin{equation}
\widehat{Z}^{\mathfrak{sl}(2|1)}\approx 
\frac{2}{\hbar}(\gamma-\log\hbar)+\frac{1}{3}
-8\sum_{n\geq 1}\frac{(2n-1)!\,\zeta(2n)^2}{(2\pi)^{4n}}\,\hbar^{2n-1},\qquad \hbar\rightarrow 0+
\label{lambert-asympt}
\end{equation}
where $\gamma$ is the Euler constant and $\zeta(2n)$ are the values of the Riemann $\zeta$-function\footnote{They can be expressed via Bernoulli numbers for even integers $2n$:
\begin{equation}
    B_{2n}=\frac{(-1)^n\,2\,(2n)!}{(2\pi)^{2n}}\,\zeta(2n).
\end{equation}
}.
The Borel transform of the non-singular part of the series reads:
\begin{equation}
	{B}(\xi):=
	\frac{1}{3}-8\sum_{n\geq 1}\frac{\zeta(2n)^2}{(2\pi)^{4n}}\,\xi^{2n-1}. 
	\label{B-series-sphere}
\end{equation}
Indeed it has a finite radius of convergence, because $\zeta(2n)\rightarrow 1$ as $n\rightarrow \infty$. Moreover, note that
\begin{equation}
	\zeta(2n)^2=\left(\sum_{m\geq 1}\frac{1}{m^{2n}}\right)^2=\sum_{m\geq 1}\frac{d(m)}{m^{2n}}.
\end{equation}
Using that
\begin{equation}
    2\sum_{n\geq 1}\left(\frac{\xi}{-4\pi^2} \right)^{2n-1}\,\frac{1}{m^{2n}}=
    \frac{1}{m-\frac{\xi}{-4\pi^2}} - \frac{1}{m+\frac{\xi}{-4\pi^2}}
\end{equation}
it follows that $B(\xi)$ has singularities at (and only at) $\xi=S_m$ where $S_m=-4\pi^2\,m$, $m\in \Z\setminus \{0\}$. The singularities are simple poles:
\begin{equation}
	B(\xi)\sim \frac{4d(|m|)}{\xi+4\pi^2 m},
	\qquad \xi\rightarrow -4\pi^2m.
	\label{B-poles-sphere}
\end{equation}
They all correspond to the same value of Chern-Simons invariant:
\begin{equation}
    S_m/(-4\pi^2)=0\mod 1,
\end{equation} 
achieved on a trivial flat connection. 

Moreover, it is possible to show that the full  $\widehat{Z}^{\mathfrak{sl}(2|1)}$ can be recovered exactly as a certain linear combination of contour integral of $B(\xi)e^{-\xi/\hbar}$. First note that $|q|$ the difference of the integrals over rays slightly shifted from the positive half of the real axis can be explicitly evaluated by the residues of the poles (\ref{B-poles-sphere}):
\begin{multline}
	-\frac{1}{2\hbar}\int_0^{(1+i\epsilon)\infty} d\xi\,B(\xi)e^{-\xi/\hbar} +\frac{1}{2\hbar}\int_0^{(1-i\epsilon)\infty} d\xi\,B(\xi)e^{-\xi/\hbar}=\\
	=\frac{4\pi i}{\hbar}\sum_{m\geq 1}d(m)e^{-\frac{4\pi^2m}{\hbar}}=
	\frac{2\pi i}{\hbar}(\widehat{Z}^{\mathfrak{sl}(2|1)}|_{\hbar\rightarrow \frac{4\pi^2}{\hbar}}+1/6)
	\label{B-integral-sphere-poles}
\end{multline}
for some small $\epsilon>0$. The exponentials $e^{-\xi/\hbar}$ in the integrals decay along the rays towards infinity because
\begin{equation}
    |q|< 1\;\Leftrightarrow \; \mathrm{Im}\,\tau >0 
    \;\Leftrightarrow \; \mathrm{Re}\,\hbar >0. 
    \label{upper-half-plane-condition}
\end{equation}
The Borel transform can be rewritten as the following integral, allowing analytic continuation:
\begin{equation}
	B(\xi)=\frac{1}{3}+\frac{i}{\pi}\int_{\text{Re}(s)=3/2} ds\,\frac{\zeta(s)^2}{\sin\frac{\pi s}{2}}
	\, \left(\frac{\xi}{4\pi^2 i}\right)^{s-1}.
\end{equation}
The sum over $n$ in (\ref{B-series-sphere}) can be recovered as the sum over the residues in the integral above.

Moreover, from the results of \cite{bettin2013reciprocity,Bettin_2013} (also cf. Section \ref{sec:q-mod-sphere}) we have  
\begin{multline}
	\widehat{Z}^{\mathfrak{sl}(2|1)}+\frac{2\pi i}{\hbar}\widehat{Z}^{\mathfrak{sl}(2|1)}|_{\hbar\rightarrow \frac{4\pi^2}{\hbar}} =\\
	\frac{1}{3}+\frac{2}{\hbar}(\gamma-\log\hbar-\pi i/6)-
	\frac{2i}{\hbar}\int_{\mathrm{Re}(s)=-1/2}\,
	\frac{\zeta(s)\zeta(1-s)}{\sin\pi s}\,
	\left(\frac{i\hbar}{2\pi}\right)^{1-s}.
	\label{Z-sphere-error}
\end{multline}
The right hand side can be understood as the error of the Eisenstein series of weight one being a weight one modular form in $\tau=\frac{i\hbar}{2\pi}$, see the next section for more details.

On the other hand, assuming in addition to (\ref{upper-half-plane-condition}) that $\mathrm{Im}\,\hbar <0$, we have
\begin{multline}
	\frac{1}{\hbar}\int_0^{-i\infty} d\xi\,\left(B(\xi)-\frac{1}{3}\right)e^{-\xi/\hbar} =
	\frac{i}{\pi}\int_{\text{Re}(s)=3/2} ds\,\frac{\zeta(s)^2\,\Gamma(s)}{\sin\frac{\pi s}{2}}\,\left(\frac{\hbar}{4\pi^2 i}\right)^{s-1}= \\
-
	\frac{2i}{\hbar}\int_{\mathrm{Re}(s)=-1/2}\,
	\frac{\zeta(s)\zeta(1-s)}{\sin\pi s}\,
	\left(\frac{i\hbar}{2\pi}\right)^{1-s}
	\label{B-sphere-error}
\end{multline}
where in the last equality we applied the standard functional equation on the Riemann $\zeta$-function and made a change of variables $s\rightarrow 1-s$.
Combining (\ref{B-poles-sphere}), (\ref{Z-sphere-error}) and (\ref{B-sphere-error}) together we than obtain the following exact expression for the homological block in terms of a linear combination of contour integrals of the Borel transform:
\begin{equation}
		\widehat{Z}^{\mathfrak{sl}(2|1)}=
\frac{2}{\hbar}(\gamma-\log\hbar) 
+\frac{1}{2\hbar}\left(\int_0^{(1+i\epsilon)\infty}+\int_0^{(1-i\epsilon)\infty}\right) d\xi\,B(\xi)e^{-\xi/\hbar}. 
\end{equation}
We used the fact that the contour going from $0$ to $-i\infty$ can be rotated to the contour going to $(1-i\epsilon)\infty$ without encountering any singularities. This gives a formula for the exact Borel resummation similar to the one in \cite{Gukov:2016njj}.

\subsection{Seifert manifolds with 3 exceptional fibers}
\label{sec:resurgence-seifert}
In this section we consider two specific examples of integer homology spheres.  This analysis is straightforward to generalize for any Seifert manifold with 3 exceptional fibers. 

\subsubsection{$\overline{\Sigma(2,3,5)}$}
\label{sec:resurgence-235}
The unique homological block $\widehat{Z}^{\mathfrak{sl}(2|1)}$ for $M^3=\overline{\Sigma(2,3,5)}$ was considered in Section \ref{sec:ex-235} and was shown to be a linear combinations of $q$-series of the form
\begin{equation}
	\sum_{m\geq 0} \frac{q^{\alpha m^2+\beta m+\gamma}}{1-q^{Am+B}}=F(q;\alpha,\beta,A,B)q^{\gamma}
	\label{q-series-5params}
\end{equation}
for some parameters $\alpha,\beta,\gamma,A,B$. The series $F(q;\alpha,\beta,A,B)$ have the overall factor $q^{\gamma}$ removed. Their asymptotic expansion is analyzed in Appendix \ref{app:asympt-F}. Note that $F(q;\alpha,\beta,A,B)$ has the same resurgence properties as the series (\ref{q-series-5params}) as its ratio is an entire function in $\hbar$ with expansion starting with $1$:
\begin{equation}
	q^\gamma\equiv e^{-\gamma \hbar}=1+O(\hbar).
\end{equation}
Namely, their Borel transforms have singularities at the same positions and, moreover, they have the same behavior near the singularities in the leading order. This implies that the Stokes jumps (\ref{stokes-jumps}) are also the same in the leading order in $\hbar$.

The terms in (\ref{PHS-q-series}) involving $P_5$ are of the form
\begin{equation}
	\sum_{m\geq 1}\frac{q^{30m(m-1)}}{1-q^m}\,q^{ma} 
	= F(q;30,30+a,1,1)\,q^{a}
\end{equation}
for some $a\in \mathbb{Z}_+$ corresponding to the powers appearing in the polynomial $P_5$. The terms in (\ref{PHS-q-series}) involving $P_1,P_2,P_3$ are of the form
\begin{multline}
	\sum_{m,n\geq 1} q^{[a+p(m-1)]((30/p)m+n)}=
	\sum_{m\geq 0} \frac{q^{30m^2+(30a/p+30+p)m}}{(1-q^{pm+n})}\,q^{a(30/p+1)}=\\
	=F(q;30,30a/p+30+p,p,a)\,q^{a(30/p+1)}
\end{multline}
with $p=15,10,6$ respectively and $a\in \Z_+$ corresponding to the powers appearing in the polynomials $P_{1},P_2,P_3$. Finally, the terms in (\ref{PHS-q-series}) involving $P_4$ are the same as the Lambert series already analyzed in Section \ref{sec:resurgence-3sphere}. The analysis of Appendix \ref{app:asympt-F} then tells us that the Borel transform of the asymptotic expansion of $\widehat{Z}^{\mathfrak{sl}(2|1)}$ at $\hbar\rightarrow 0$ has possible singularities at $\xi = S$ with
$S/(-4\pi^2)=-K^2/120$  and $S/(-4\pi^2)=(2K(30/p))^2/120$ with $30/p=2,3,5,30$ and $K\in \mathbb{Z}_+$. The corresponding Stokes jumps are given in the leading order by equations (\ref{F-stokes-jump-type-I}) and (\ref{F-stokes-jump-type-II}) respectively. However, for  most of the values of $S$  the contributions from different terms in (\ref{237-q-series}) actually cancel out in a non-trivial way, at least in the leading order. The table below lists the total values of the Stokes jumps combined from all the terms in (\ref{PHS-q-series}) for a few small values of $S$.
\begin{equation}
\begin{tabular}{c|c}
    $\frac{S}{-4\pi^2}$ & $\frac{\Delta Z}{2i\sqrt{\pi}}\,\hbar^{1/2}\,e^{\frac{S}{\hbar}}\mod O(\hbar)$ \\
    \hline
     $-\frac{1}{120}$ & $-2\sqrt{2+4/\sqrt{5}}$  \\
     $-\frac{4}{120}$ &  0 \\
     $-\frac{9}{120}$ &  0 \\
     $-\frac{25}{120}$ &  0 \\
      $-\frac{49}{120}$ &  $-2\sqrt{2-4/\sqrt{5}}$ \\
     $\cdots$ & $\cdots$ \\
 \end{tabular}
\qquad
 \begin{tabular}{c|c}
    $\frac{S}{-4\pi^2}$ & $\frac{\Delta Z}{2\pi i}\,\hbar\,e^{\frac{S}{\hbar}}\mod O(\hbar)$ \\
    \hline
     $\frac{16}{120}$ & 0  \\
     $\frac{36}{120}$ & 0  \\
    $\cdots$ & $\cdots$ \\
 \end{tabular}
\end{equation}
Note that the values $S/(-4\pi^2)=-16/120$ and $S/(-4\pi^2)=-36/120$ are skipped because the formula (\ref{cL-as}) is inapplicable when $KA/\alpha \in 2\mathbb{Z}$ and in particular when $(K,A,\alpha)=(4,15,30)$ or $(6,10,30)$.

The results are in agreement with the critical values of the Chern-Simons action functional of $SU(2|1)$ flat connections on the Poincare homology sphere $\overline{\Sigma(2,3,5)}$. As was explained above, when $H^1(M^3,\Z)\cong 0$ one should compare the positions of the singularities of the Borel transform with the values of the Chern-Simons invariant of $SL(2,\C)$ flat connections on $M^3$. Its non-abelian flat connections can be conjugated to the $SU(2)$ subgroup. It is known that there are two non-abelian $SU(2)$ flat connections on $\Sigma(2,3,5)$ for which the values of the Chern-Simons functional  are $-1/120$ and $-49/120$ modulo $1$. There is only one abelian flat connection: the trivial flat connection.

\subsubsection{${\Sigma(2,3,7)}$}
\label{sec:resurgence-237}

The resurgence analysis for $M^3={\Sigma(2,3,7)}$ is quite similar to the case of $M^3=\overline{\Sigma(2,3,5)}$ considered above, however this example will have a new feature. The terms in (\ref{237-q-series}) involving $P_5$ are of the form
\begin{equation}
	\sum_{m\geq 1}\frac{q^{42m(m-1)}}{1-q^m}\,q^{ma} 
	= F(q;42,42+a,1,1)\,q^{a}
\end{equation}
for some $a\in \mathbb{Z}_+$ corresponding to the powers appearing in the polynomial $P_5$. The terms in (\ref{PHS-q-series}) involving $P_1,P_2,P_3$ are of the form
\begin{multline}
	\sum_{m,n\geq 1} q^{[a+p(m-1)]((42/p)m+n)}=
	\sum_{m\geq 0} \frac{q^{42m^2+(42a/p+42+p)m}}{(1-q^{pm+n})}\,q^{a(42/p+1)}=\\
	=F(q;42,42a/p+42+p,p,a)\,q^{a(42/p+1)}
\end{multline}
with $p=21,14,6$ respectively and $a\in \Z_+$ corresponds to the powers appearing in the polynomials $P_{1,2,3}$. The analysis of Appendix \ref{app:asympt-F} tells us that the Borel transform of the asymptotic expansion of $\widehat{Z}^{\mathfrak{sl}(2|1)}$ at $\hbar\rightarrow 0$ has possible singularities at $\xi = S$ with
$S/(-4\pi^2)=-K^2/168$ and $S/(-4\pi^2)=(2K(42/p))^2/168$ with $42/p=2,3,7,42$ and $K\in \mathbb{Z}_+$. However, as in the case of $M^3=\overline{\Sigma(2,3,5)}$ for most of the values of $S$ the contributions from different terms in (\ref{237-q-series}) cancel out in a non-trivial way, at least in the leading order. The table below lists the total values of the Stokes jumps combined from all the terms in (\ref{237-q-series}) for a few small values of $S$.
\begin{equation}
\begin{tabular}{c|c}
     $\frac{S}{-4\pi^2}$ & $\frac{\Delta Z}{2i\sqrt{\pi}}\,\hbar^{1/2}\,e^{\frac{S}{\hbar}}\mod O(\hbar)$ \\    \hline
     $-\frac{1}{168}$ & $\sqrt\frac{8}{7}\,\tan \frac{3\pi}{8}$  \\
     $-\frac{4}{168}$ &  0 \\
     $-\frac{9}{168}$ &  0 \\
     $-\frac{25}{168}$ &  $-\sqrt\frac{8}{7}\,\tan \frac{\pi}{8}$ \\
      $-\frac{49}{168}$ & 0 \\
    $-\frac{81}{168}$ & 0 \\
    $-\frac{100}{168}$ & 0 \\
      $-\frac{121}{168}$ & $-\sqrt\frac{8}{7}\,\tan \frac{2\pi}{8}$  \\
       $-\frac{169}{168}$ & $-\sqrt\frac{8}{7}\,\tan \frac{3\pi}{8}$\\
       $\cdots$ & $\cdots$ \\
 \end{tabular}
 \qquad
 \begin{tabular}{c|c}
   $\frac{S}{-4\pi^2}$ & $\frac{\Delta Z}{2\pi i}\,\hbar\,e^{\frac{S}{\hbar}}\mod O(\hbar)$ \\   \hline
    $\frac{16}{168}$ & 0  \\
     $\frac{36}{120}$ & 0  \\
      $\frac{64}{120}$ & 0  \\
        $\frac{121}{120}$ & 0  \\
        $\cdots$ & $\cdots$ \\
 \end{tabular}
\end{equation}

Note that the values $S/(-4\pi^2)=-16/168,-36/168,-64/168,-144/168,\ldots$ are skipped as the formula (\ref{cL-as}) is inapplicable when $KA/\alpha \in 2\mathbb{Z}$ and in particular when $(K,A,\alpha)=(4,21,42),(6,14,42),(8,21,42),(12,14,42),\ldots$.

The results are again in agreement with the critical values of the Chern-Simons action. As in the previous case, $H^1(M^3,\Z)\cong 0$ and one should compare the positions of the singularities of the Borel transform with the values of the Chern-Simons invariant of $SL(2,\C)$ flat connections. There are $2$ non-abelian $SL(2,\mathbb{C})$ flat connections on $\Sigma(2,3,7)$ that can be conjugated into the $SU(2)$ subgroup and their values of the Chern-Simons functional are $-25/168$ and $-121/168$ modulo $1$. There is also a single non-abelian $SL(2,\mathbb{C})$ flat connection that can be conjugated into the $SL(2,\mathbb{R})$ subgroup and the corresponding value of the Chern-Simons functional is $-1/168$. Finally, there is a single abelian flat connection, which is the trivial one.

\section{Quantum modularity}
\label{sec:q-mod}
The resurgence analysis of the previous section opens the door to the study of the modular structure of the homological blocks. Interestingly, the transformation properties of $\widehat{Z}^{\mathfrak{sl}(2|1)}[S^3]$ analyzed in \ref{sec:resurgence-3sphere} and the ones of $\widehat{Z}^{\mathfrak{sl}(2|1)}[L(p,1)]$, that we will examine later in the section, hint at a connection to quantum modular forms. 

As opposed to modular forms, quantum modular forms are defined at the boundary of the upper half-plane, they do not transform covariantly under the action of the modular group, and moreover they are neither analytic nor continuous functions\footnote{They are defined on $\mathbb{P}^1(\Q)$, endowed with the discrete topology.}. 
Nonetheless, the action of the modular group on these functions reveals some of their interesting properties: the obstruction to modularity of a quantum modular form is a ``nicer'' (``more analytic'') function than the quantum modular form itself. 
In other words, a quantum modular form of weight $w$ on $SL_2(\Z)$ is a function $Q: \Q \rightarrow \C$ such that for every $\gamma=(a\;b;\,c\;d)\in SL_2(\Z)$ the function $p_{Q,\gamma} (x): \Q \setminus \{{\gamma}^{-1}(\infty)\} \rightarrow \C$ defined by 
\begin{equation}
    \label{eq:qum}
p_{Q,\gamma}(x):=Q(x) - (cx+d)^{-w}Q(\frac{ax+b}{cx+d})
\end{equation}
is a better behaved function. The function $\gamma \mapsto p_{Q,\gamma}$ is a cocycle on $SL_2(\Z)$. 
See \cite{ZagQMF} for the original definition and various examples. 

The analytic properties of the modular obstruction $p_{Q,\gamma}(x)$ allow us to differentiate between different types of quantum modular forms. 
In the following, we will consider examples of what are usually known as strong quantum modular forms.

When $Q(x)$ is a strong quantum modular form, not only $p_{Q,\gamma}(x)$ is a real-analytic function on $\mathbb{P}^1(\R)$ (possibly minus a finite subset) but one can also associate to $Q(x)$ a formal power series so that \eqref{eq:qum} holds as an identity between (countable collections of) formal power series.

Many examples of strong quantum modular forms arise as the radial limit of holomorphic functions on the upper half-plane $\mathcal{H}$, which do not transform nicely under a modular transformation but those obstruction to modularity is measured by $p_{Q,\gamma}(\tau)$. The modular structure of these holomorphic functions naturally leads to the definition of a quantum modular form on the boundary of $\mathcal{H}$.

Mock and false theta functions are a particular example of the above  \cite{LawZag}. 
Interestingly, in the context of 3-manifold invariants, they appear as $\mathfrak{sl}(2)$ homological blocks of certain rational homology spheres \cite{Cheng:2018vpl,bringmann2018quantum, Cheng_2019} and their quantum modular properties are reflected in the WRT invariants \cite{LawZag, Hikami_2005}.

In the rest of this section, we investigate yet another type of strong quantum modular forms that appears in connection to $\widehat{Z}^{\mathfrak{sl}(2|1)}[L(p,1)]$ and turns out to be related to the Eisenstein series of weight one. 
One of the interesting features of these objects is that the period function\footnote{Here we consider the case when $\gamma$ is $S=\biggl( \begin{matrix}0&-1\\ 1&0\end{matrix} \biggr)$.} $p_{Q,S}$ extends to a function on the complex plane minus the negative real axis.

\subsection{3-sphere}
\label{sec:q-mod-sphere}
From Section \ref{sec:ex-sphere} we know that the $\mathfrak{sl}(2|1)$ homological block of the $S^3$ reads 
	 \begin{equation}
		 \widehat{Z}^{\mathfrak{sl}(2|1)}(\tau)=-\frac{1}{6}+2\sum_{n\geq 1}\frac{q^n}{1-q^n}=\frac{1}{3} + 2G_1(\tau)
	\end{equation}
where $G_1(\tau)$ is the weight 1 holomorphic Eisenstein series defined in equation \eqref{eq:eisentein1} and $d(m)$ is the number of divisors of $m$. 

Denote by $F(\tau)$ the Lambert series 
\begin{equation}
\label{eq:F}
F(\tau):= \sum_{n\geq 1}\frac{q^n}{1-q^n}.
\end{equation} 
To this holomorphic and periodic function one can associate the period function\footnote{To shorten the notation, in the following we take $p_{Q}(\tau):= p_{Q,S}(\tau)$.} \cite{bettin2013reciprocity, Bettin_2013}
\begin{equation}
\label{eq:periodF}
p_{F}(\tau):= F(\tau)- \frac{1}{\tau}F(-1/\tau),
\end{equation}
a real analytic function with growth $p_{F}(\tau)=O(|\text{log}(\tau)|/\tau)$ as $\tau \rightarrow 0^+$ and $p_{F}(\tau)=O(\text{log}(\tau))$ as $\tau\rightarrow \infty$ and for which the three term relations
\begin{equation}
p_{F}(\tau)=p_{F}(\tau+1)+(\tau+1)^{-1}p_{F}(\frac{\tau}{\tau+1})
\end{equation}
holds.

As proven in  \cite{bettin2013reciprocity}, the period function, initially defined in the upper half-plane, can be analytically continued to the slit plane $\mathbb{C}':=\mathbb{C}\backslash \mathbb{R}_{\le 0}$
via the representation
\begin{equation}
\label{eq:pF}
p_{F}(\tau)= r(\tau)+ \frac{1}{2\pi}\int_{\text{Re}(s)=-1/2} ds\, (2\pi \tau)^{-s} \frac{\Gamma(s) \zeta(s)^2}{\text{sin}\bigl(\frac{s\pi}{2}\bigr)}
\end{equation}
where
\begin{align}
r(\tau) &= \frac{\text{log}(-2\pi i\tau) -\gamma}{2\pi i \tau} + \frac{1}{4}.
\end{align}
Up to a non-smooth term, the radial limit of $F(\tau)$ provides examples of quantum modular forms \cite{bettin2013reciprocity}.

\subsection{Lens spaces}
\label{sec:q-mod-lens}

Consider now the expression of the $\mathfrak{sl}(2|1)$ homological blocks of $L(p,1)$ given in \eqref{Zhat-lens-explicit}.
The homological blocks of the lens space $L(2,1)$ can be proven to satisfy the following transformation properties under the action of the generators of $SL_2(\Z)$, 
\begin{align}
\label{eq:q-mod-L21-S}
&\widehat{Z}^{\mathfrak{sl}(2|1)}_{a,b}(\tau) - \frac{1}{2\tau} \sum_{a',b'}(-1)^{aa'+bb'}\widehat{Z}^{\mathfrak{sl}(2|1)}_{a',b'}(-1/\tau) = \psi_{2,(a,b)}(\tau)\\
\label{eq:q-mod-L21-T}
&\widehat{Z}^{\mathfrak{sl}(2|1)}_{a,b}(\tau) = (-1)^{ab} \widehat{Z}^{\mathfrak{sl}(2|1)}_{a,b}(\tau+1)
\end{align}

The function $\psi_{2,(a,b)}(\tau)$ extends to an analytic function in the slit plane $\C'$ and it is given by
\begin{align}
    \psi_{2,(a,b)}(\tau) &=\widetilde{r}_{2,(a,b)} + \frac{1}{\pi}\int_{\text{Re}(s)=-1/2} ds\, (2\pi \tau)^{-s}  \frac{\Gamma(s)}{\text{sin}\bigl(\frac{s\pi}{2}\bigr)}2^{-s}\zeta(s,a/2)\zeta(s,b/2),
    \end{align}
    with 
    \begin{align}
\widetilde{r}_{2,(a,b)}(\tau) &= 
\text{const}_{a,b}+ \frac{1}{2}(a-1)(b-1)+\frac{1}{2\pi i \tau}{\times}\notag\\
& {\times}\bigl(\text{log}(-4\pi i\tau)+\gamma-\gamma_0(a/2)-\gamma_0(b/2){-} \pi i\sum_{a',b'}(-1)^{aa'+bb'}\text{const}_{a',b'}\bigr)
\end{align}
where $\zeta(s,q)$ and $\gamma_0(s)=-\psi(g)$ are the Hurwitz zeta function and minus the digamma function respectively (see Appendix \ref{app:zeta} for the definitions).
The $\mathfrak{sl}(2|1)$ homological blocks of $L(2,1)$, thus provide, up to a non-smooth correction term, new examples of vector-valued quantum modular forms when approaching the real line. 

In Appendix \ref{app:zeta} we give two separate proofs of the above: the first one is based on the relation between $\widehat{Z}^{\mathfrak{sl}(2|1)}_{a,b}(\tau)$ and the Lambert series $F(\tau)$, while the second one follows the method described in \cite{Bettin_2013, Lewis_2019}.

For more general lens spaces we expect a similar behavior under the action of the modular group. More specifically, we conjecture that
\begin{align}
&\widehat{Z}^{\mathfrak{sl}(2|1)}_{a,b}(\tau) - \frac{1}{p\tau} \sum_{a',b'}e^{2\pi i\frac{aa'+bb'}{p}}\widehat{Z}^{\mathfrak{sl}(2|1)}_{a',b'}(-1/\tau) = \psi_{p,(a,b)}(\tau)\\
&\widehat{Z}^{\mathfrak{sl}(2|1)}_{a,b}(\tau) = e^{2\pi i\frac{ab}{p}} \widehat{Z}^{\mathfrak{sl}(2|1)}_{a,b}(\tau+1)
\end{align}
where $\psi_{p,(a,b)}(\tau)$ is given by
\begin{align}
    \psi_{p,(a,b)}(\tau) &=\widetilde{r}_{p,(a,b)} + \frac{1}{\pi}\int_{\text{Re}(s)=-1/2} ds\, (2\pi \tau)^{-s}  \frac{\Gamma(s)}{\text{sin}\bigl(\frac{s\pi}{2}\bigr)}p^{-s}\zeta(s,a/p)\zeta(s,b/p),
    \end{align}
    where the integral converges for all $\tau\in \C$ with $|\text{arg}(\tau)|<\pi$ and thus it provides an extension of $\psi$ to the slit plane. The function $\widetilde{r}_{p,(a,b)}$ is
    \begin{align}
\widetilde{r}_{p,(a,b)}(\tau) &= 
\text{const}_{a,b}+ 2\zeta(0,a/p)\zeta(0,b/p)+\frac{1}{\pi i p\tau}{\times}\notag\\
&{\times} \bigl(\text{log}(-2\pi ip\tau)+\gamma-\gamma_0(a/p)-\gamma_0(b/p){-}\pi i\sum_{a',b'}(-1)^{aa'+bb'}\text{const}_{a',b'}\bigr)\,.
\end{align}

\section{Comparison with invariants from unrolled quantum supergroups}

\label{sec:quantum-supergroup}
	The standard Witten-Reshetikhin-Turaev approach to topological invariants of links and 3-manifolds applied for quantum $\mathfrak{sl}(2|1)$ leads to trivial or ill-defined invariants as, in particular, quantum (super)dimensions vanish. From the perspective of (physical) path integrals this issue is addressed in \cite{Mikhaylov:2014aoa}. From a mathematical point of view, the issue of vanishing quantum dimensions can be overcome by introducing the notion of modified quantum dimensions \cite{geer2009modified,geer2011generalized}. 
	 Through this concept one can construct non-trival invariants of links associated to $\mathfrak{sl}(2|1)$ \cite{geer2007multivariable} that, in a certain sense, generalize Links-Gould invariants \cite{links1992two}. The introduction of the modified dimensions comes at the cost of coloring the link components by representations that are labelled by continuous parameters and form a non-semisimple category. This leads to an obvious obstruction to extending the link invariants to invariants of 3-manifolds, as the input of the standard Reshetikhin-Turaev construction is a modular tensor category, which, in particular, has finite number of simple objects. This problem was in turn overcame in \cite{costantino2014quantum} where the authors constructed quantum invariants of 3-manifolds associated to a non-semisimple category of certain representations of $\mathcal{U}^H_q(\mathfrak{sl}(2))$, the unrolled version of quantum $\mathfrak{sl}(2)$, at roots of unity.
	 A similar construction was more recently applied in \cite{ha2018topological} to the non-semisimple category of representations (satisfying certain conditions) of
	 $\mathcal{U}^H_q(\mathfrak{sl}(2|1))$, the unrolled version of quantum $\mathfrak{sl}(2|1)$ at\footnote{In \cite{ha2018topological} the quantum deformation parameter is denoted by $\xi$. The latter is related to the variable $q$ (as used in this paper) as $\xi^2=q$.} $q=e^{\frac{4\pi i}{\ell}}$, for odd $\ell$. Here we provide only a very brief review of the key elements of the construction and for the details we refer to the original paper. The result is an invariant\footnote{In \cite{ha2018topological} a more general invariant $N_\ell(M^3,T,\omega)$ is considered, where $T$ is a colored framed link inside $M^3$. For our purposes  we consider the case $T=\emptyset$.} 
	\begin{equation}
		N_\ell(M^3,\omega) 
	\end{equation}
	of closed oriented connected 3-manifolds $M^3$ ``colored'' by 
	\begin{equation}
\omega\in \mathcal{C}:=H^1(M^3,C)\setminus \bigcup_{i=1}^{3} H^1(M^3,C_i)
\label{CGP-omega}
 \end{equation}
where $C_i$ are the following subgroups of $C:=\C/\Z\times \C/\Z$:
\begin{equation}
    \begin{array}{rcl}
         C_1 & = & \{(X,Y)\in \C/\Z\times \C/\Z \;|\;
         2X=0\mod 1\}, \\
         C_2 & = & \{(X,Y)\in \C/\Z\times \C/\Z \;|\;
         2Y=0\mod 1\}, \\
         C_3 & = & \{(X,Y)\in \C/\Z\times \C/\Z \;|\;
         2(X+Y)=0\mod 1\}. \\
    \end{array}
\end{equation}
Note that in particular $\mathcal{C}$ is empty for integer homology spheres and, more generally, for manifolds such that all elements of $H^1(M^3,\Z)$ have order at most 2. For a given 3-manifolds, the values of the invariant are determined from its  realization by a surgery on a framed link in $S^3$. Without going into details, in order to compute the invariant, one starts with the corresponding invariant of the framed link with each component $I$ labeled by the elements $(X_I,Y_I)\in \C\times \C$ such that 
\begin{equation}
    2X_I\notin \Z,\;2Y_I\notin\Z,\;2(X_I+Y_I)\notin\Z.
    \label{CGP-rep-conditions}
\end{equation}
These labels correspond to highest weights of certain representations of $\mathcal{U}^H_q(\mathfrak{sl}(2|1))$. To reproduce the invariant of the 3-manifold, for each link component $I$ one then sums over a finite subset of representations with weights satisfying
\begin{equation}
    (X_I,Y_I)\mod \Z\times \Z=\omega(\mathfrak{m}_I)
\end{equation}
where $\mathfrak{m}_I$ is an element of $H_1(M^3,C)$ represented by a meridian on the boundary of the tubular neighborhood of the link component\footnote{That is a small cycle going once around the link component.}. Note that for an arbitrary surgery representation of $M^3$ and an element (\ref{CGP-omega}) it is not always true that 
that $\omega(\mathfrak{m}_I)\in C\setminus \cup_{i=1}^3 C_I$ and the weights $(X_I,Y_I)$ can satisfy the conditions (\ref{CGP-rep-conditions}). If it is true that  $\omega(\mathfrak{m}_I)\in C\setminus \cup_{i=1}^3 C_I$ for all link components $I$, the surgery representation is called \textit{computable} for a given pair $(M^3,\omega)$.

In Appendix \ref{app:gauss-sums} we consider the invariant $N_\ell(M^3,\omega)$ in the case when $M^3$ is a general plumbed 3-manifold of the type considered in the previous sections. Using a generalized Gauss reciprocity formula we obtain its analytic continuation with respect to $q=e^{\frac{4\pi i}{\ell}}$. We then rewrite the result 
in the universal terms, which leads to a more general conjecture about relation between the invariant $N_\ell(M^3,\omega)$ and $\widehat{Z}_{a,b}^{\mathfrak{sl}(2|1)}[M^3]$ in the radial limit $q\rightarrow e^{\frac{4\pi i}{\ell}}$, which we formulate below. 

As before, consider the case when $M^3$ is a rational homology sphere, that is $H_1(M^3,\Z)$ is a finite abelian group. This implies that $H^1(M^3,\C/\Z)\cong H^1(M^3,\Q/\Z)$ and therefore the two components of (\ref{CGP-omega}) can be understood as the elements of $H^1(M^3,\Q/\Z)$:
\begin{equation}
    \omega=(\omega_1,\omega_2),\;\omega_i\in H^1(M^3,\Q/\Z),\;2\omega_i\neq 0,\;2(\omega_1+\omega_2)\neq 0.
\end{equation}
Moreover, as $\mathrm{Hom}(H_1(M^3,\Z),\Q/\Z)\cong H^1(M^3,\Q/\Z)$, the linking pairing (\ref{linking-pairing-def}) provides an isomorphism $H^1(M,\Q/\Z)\cong H_1(M^1,\Z)$ for a rational homology sphere. Assuming these identifications, we have the following conjecture for a rational homology $M^3$:
	\begin{multline}
	    N_\ell(M^3,\omega) =
	   \frac{\pm\mathcal{T}([2\omega_1])}{\ell|H_1(M^3,\Z)|}\times \\
	   \times\sum_{\substack{\beta,\gamma \in H_1(M^3,\Z) \\ b,c\in H^1(M^3,\C/\Z)}}
	e^{2\pi i k \cdot \ell k(\beta,\gamma)+4\pi i(b-\omega_2)(\gamma)+2\pi i(c-(\omega_1+\omega_2))(\beta)}    \cdot \widehat{Z}^{\mathfrak{sl}(2|1)}_{b,c}|_{q\rightarrow e^{\frac{4\pi i}{\ell}}}
	\end{multline}
	where $\mathcal{T}([2\omega_1])$ is the Reidemeister torsion (equal to the analytic torsion) of the $U(1)$ flat connection specified by
	\begin{equation}
	    [2\omega_1]:=\left(2\omega_1\mod H^1(M^3,\Z)\right) \; \in H^1(M^3,\Q/\Z),
	\end{equation}
same as appeared in the context of $U(1|1)$ Chern-Simons theory \cite{ray1971r,muller1978analytic,cheeger1977analytic,Rozansky:1992zt,nicolaescu2008reidemeister,Mikhaylov:2015nsa}. The overall sign $\pm$ is due to the intrinsic sign ambiguity of the torsion. The ambiguity can be fixed by introducing some additional structure, such as an Euler structure or a $\text{spin}^c$ structure \cite{turaev1990euler,turaev1997torsion}, or a spin structure \cite{Mikhaylov:2015nsa}. 
We do not address fixing the sign ambiguity in this work. In addition, note that in the case of $\mathfrak{g}=\mathfrak{sl}(2)$, the connection between $\widehat{Z}^{\mathfrak{g}}$  and the quantum invariants associated to a non-semisimple category of representations of $\mathcal{U}^H_q(\mathfrak{sl}(2))$ was discussed in \cite{Gukov:2020lqm}, where a conjectural formula relating the corresponding invariants of  complements of knots in $S^3$ was given.

\section{Some open questions}
\label{sec:open-questions}

In this final section we give a brief list of some open questions which  lie beyond the scope of this paper. Some of these questions have already appeared earlier in the text.

\begin{itemize}
    \item The integration contour in the formula for $\widehat{Z}^{\mathfrak{sl}(2|1)}[M^3]$ for a plumbed $M^3$ was (uniquely) fixed by the requirement that the result is a well defined $q$-series. First, it would be interesting to provide a more physical interpretation of this choice of contour, or, equivalently, the choice of the expansion chamber of the integrand. Second of all, in order to define the invariant for the oppositely oriented 3-manifold, $\overline{M^3}$, one would need to understand how to regularize the integral such that it gives a nice $q$-series when the inequalities in  \eqref{plumbing-cond-1}- \eqref{plumbing-cond-3} are reversed. Equivalently, it would be interesting to understand how to extend the homological blocks outside their region of convergence.

\item It is of interest to understand for which other supergroups, other than $SU(2|1)$, the contour integral expression (\ref{Zhat-plumbed-superg}) for $\widehat{Z}^{\mathfrak{g}}$ of plumbed 3-manifold can be given a precise meaning, and fix the choice of the contour, or equivalently, the expansion chamber of the integrand.

\item In Section \ref{sec:lens} a relation between $\widehat{Z}^{\mathfrak{sl}(2|1)}[L(p,1)]$ and counting of $SO(3)\cong SU(2)/\Z_2$ instantons on $L(p,1)\times \mathbb{R}$ was noticed. It would be interesting to explore if similar relation holds for 3-manifolds other than $L(p,1)$. Note that from the brane realization of $\widehat{Z}^{\mathfrak{g}}$ considered in Section \ref{sec:brane-setup} it is actually more natural to expect a relation between counting of $SU(N)/\Z_N$ on $M^3$ and $\widehat{Z}^{\mathfrak{psl}(N|N)}[M^3]$. In the case when $N=2$ and $M^3=L(p,1)$ it is easy to show that $\widehat{Z}^{\mathfrak{psl}(2|2)}_{b,c}[L(p,1)]$ is  proportional to the derivative of $\widehat{Z}^{\mathfrak{sl}(2|1)}_{b,c}[L(p,1)]$ with respect to $\tau=\log q/(2\pi i)$, and thus essentially contains the same information.

\item Numerical experiments show that at least for all relatively simple examples of 3-manifolds, at least the first few coefficients of the $q$-series $\widehat{Z}^{\mathfrak{sl}(2|1)}$ are positive, except possibly the constant term (which can actually be interpreted as $\zeta$-regularized infinite sum of positive integers). A priory, this is not obvious from the contour integral expression. It would be interesting to explore whether such positivity holds for all the coefficients and all plumbed 3-manifolds and, if yes, understand why this occurs.

\item Of course, it would be interesting to provide a categorification of $\widehat{Z}^{\mathfrak{sl}(2|1)}[M^3]$. One can hope that at least in the case of plumbed 3-manifolds one can give an explicit description of the underlying doubly graded vector spaces, similar to the one given in \cite{ozsvath2003floer,nemethi2005ozsvath} for $\mathfrak{g}=\mathfrak{gl}(1|1)$ case.  There are various indications that a categorification in the case of $\mathfrak{g}=\mathfrak{sl}(2|1)$ might be simpler than in the case of $\mathfrak{g}=\mathfrak{sl}(2)$. In the case of plumbed 3-manifolds one can also try to find a refinement of the $q$-series by another variable, say $t$ that would correspond to an additional grading on the $Q$-homology (i.e. an extra $U(1)$ symmetry, in more physical terms). There are some indications that such a refinement can be provided by simply changing $y_I\rightarrow y_I\,t^{\alpha_I}$, $z_I\rightarrow z_I\,t^{-\alpha_I}$ in the product over the vertices in the first line of (\ref{Zhat-plumbed-sl21-CT}), and the result is still invariant under Kirby-Neumann moves. However we are not aware of an a priori reason for this quantity to be a 3-manifold invariant.

\item It would be interesting to generalize the invariants $\widehat{Z}^{\mathfrak{sl}(2|1)}$ to 3-manifolds with torus boundaries, as it was done in \cite{Gukov:2019mnk} for $\widehat{Z}^{\mathfrak{sl}(2)}$. This would allow calculating $\widehat{Z}^{\mathfrak{sl}(2|1)}$ for 3-manifolds other than plumbings, via surgery. In particular, for a complement of a knot $K$, one can expect the corresponding invariant to be valued in series in 3 variables with integer coefficients:
\begin{equation}
    F_K^{\mathfrak{sl}(2|1)}(y,z)\in \Z[[q,y^{\pm 1},z^{\pm 1}]]. 
\end{equation}
Then, for example, for a 3-manifold $M^3=S^3(K_p)$ obtained by an integral $p$-surgery (assume $-p\in \Z_+$) on $K$ we expect that the following holds:
\begin{multline}
    \widehat{Z}^{\mathfrak{sl}(2|1)}_{a,b}[S^3(K_p)]=\\
	\mathrm{CT}_{y,z}\left\{
	\left.
	\frac{y-z}{(1-y)(1-z)}
	\right|_{\substack{|y|<1\\|z|>1}}
	\cdot
	F_K^{\mathfrak{sl}(2|1)}(y,z)
	\sum_{\substack{n\in p\Z+a \\ m\in p\Z+b}}
	q^{\frac{nm}{p}}\,z^{m}y^{n} 
	\right\} \;\;\; \in  q^{\Delta_{ab}}\Z[[q]]
	\label{Zhat-surgery-sl21-CT}
\end{multline}

    \item In the case of ordinary Lie algebra, the (unreduced) homological blocks for 3- and 4-star shaped graphs coincide with the characters of certain logarithmic vertex operator algebras \cite{Cheng:2018vpl, CCFFGH}. Remarkably, the (unreduced) ${\mathfrak{sl}(2|1)}$ homological blocks of lens spaces and 3-star shaped Seifert manifolds can also be written in terms of characters of atypical modules $\widehat{\mathcal{A}}_{\ell,n}$ associated to the affine superalgebra $\widehat{\mathfrak{gl}}(1|1)$, as explained in Section \ref{sec:lens} and   \ref{sec:ex-seifert}. It would be illuminating to understand in this case the connection to this superalgebra.

\item From the resurgence analysis of Section \ref{sec:resurgence-seifert} the  structure of the singularities of the Borel transform of  $\widehat{Z}^{\mathfrak{sl}(2)}[M_3]$, where $M_3$ is a Seifert manifold with 3-singular fibers, appears to be much more complicated than the ones of $\widehat{Z}^{\mathfrak{sl}(2)}[L(p,1)]$. It would be interesting to analyze the transformation properties of $\widehat{Z}^{\mathfrak{sl}(2)}[M_3]$ under the modular group.

    \item It would be interesting to find a concrete interpretation of the mathematical invariants $N_\ell(M^3,\omega)$, considered in Section \ref{sec:quantum-supergroup}, in terms of path integral of supergroup Chern-Simons theory, where the 3-manifold ``color'' $\omega$ plays the role of a background field, similarly to the case of $U(1|1)$ gauge group, as was described in \cite{Mikhaylov:2015nsa}. 
 We hope that the resurgence analysis performed in Section \ref{sec:resurgence} may give some insight.\\
Note that a similar 3-manifold invariant, associated to the non-semisimple category of representations of the unrolled quantum supergroup $\mathcal{U}^H_{q}(\mathfrak{sl}(2))$ \cite{costantino2014quantum} is already lacking a similar interpretation, to our knowledge. Moreover, it would be interesting to explore possible quantum modular properties of the invariants $N_\ell(M^3,\omega)$ for more general 3-manifolds. 
This question is a part of a more general question: Are quantum invariants of 3-manifold always examples of quantum modular forms?

\end{itemize}

\section*{Acknowledgements}

We would like to thank Mina Aganagic, Thomas Creutzig, Sergei Gukov, Sarah Harrison, Marcos Mari\~no, Ramadevi Pichai, Edward Witten, Don Zagier for useful discussions and comments on the draft. The work of F.F. is supported in part by the MIUR-SIR grant RBSI1471GJ ``Quantum Field Theories at Strong Coupling: Exact Computations and Applications". P.P. would like to thank Institut Mittag-Leffler for hospitality during the final stage of the project.

\appendix 

\section{Choice of chamber for $\mathfrak{sl}(2|1)$}
\label{app:sl21-contour}
In this section we explain how the conditions (\ref{plumbing-cond-1})-(\ref{plumbing-cond-3}) on the existence of a good chamber arise. 

Note that the factors in (\ref{sl21-vert-prod}) corresponding to vertices of degree $\deg(I)=2$ are trivial, and therefore can be ignored.

Factors in (\ref{sl21-vert-prod}) corresponding to vertices of degree $\deg(I)=2+K>2$ can be expanded in two different ways:
\begin{multline}
\left(\frac{(1-z_I)(1-y_I)}{y_I-z_I}\right)^K
	=\\
	\left\{
	\begin{array}{ll}
	(z_I-1)^K(1-y_I^{-1})^K\sum_{\ell_I=0}^\infty \frac{(\ell_I+1)(\ell_I+2)\cdots (\ell_I+K-1)}{(K-1)!}\,z_I^{\ell_I} y_I^{-\ell_I},
	& |z_I|<|y_I|, \\
	\;\;(1-y_I)^K(1-z_I^{-1})^K\sum_{\ell_I=0}^\infty \frac{(\ell_I+1)(\ell_I+2)\cdots (\ell_I+K-1)}{(K-1)!}\,y_I^{\ell_I} z_I^{-\ell_I},
	& |z_I|>|y_I|. 
	\end{array} 
	\right.
	\label{vertex-series-high-degree}
\end{multline}
Factors in (\ref{sl21-vert-prod}) corresponding to vertices of degree $\deg(I)=1$ can be expanded in four different ways:
\begin{multline}
	\frac{y_I-z_I}{(1-z_I)(1-y_I)}
	=
	\frac{y_I}{1-y_I}-\frac{z_I}{1-z_I}= \\
	\left\{
	\begin{array}{ll}
	(a)\;\;\sum_{\ell_I=1}^\infty y_I^{\ell_I} -\sum_{\ell_I=1}^\infty z_I^{\ell_I}, 
	& |y_I|<1,\,|z_I|<1, 
	\\
	(b)\;\;\;\sum_{\ell_I=0}^\infty z_I^{-\ell_I} -\sum_{\ell_I=0}^\infty y_I^{-\ell_I}, 
	& |y_I|>1,\,|z_I|>1, 
	\\
	(c)\;\;\;\sum_{\ell_I=1}^\infty y_I^{\ell_I} +\sum_{\ell_I=0}^\infty z_I^{-\ell_I}, 
	& |y_I|<1,\,|z_I|>1, 
	\\
	(d)\;\;\;-\sum_{\ell_I=0}^\infty y_I^{-\ell_I} -\sum_{\ell_I=1}^\infty z_I^{\ell_I}, 
	& |y_I|>1,\,|z_I|<1.
	\end{array}
	\right.
	\label{vertex-series-degree-1}
\end{multline}
First, we argue that the first two options are not allowed in a good chamber for a generic plumbing, in the terminology of Section \ref{sec:Zhat-sl21}. Let $U\subset \vert|_{d=1}$ be the set of vertices of degree one for which we have expansion $(a)$ or $(b)$ and assume that it is non-empty set in a given chamber. Let $W\subset \vert|_{d\neq 2}$ be the subset of vertices of degree one  for which expansion $(c)$ or $(d)$ holds, together with all vertices of degree greater than two. Then ${V}|_{\deg \neq 2}=U\sqcup W$ and if the plumbing is generic, by definition, there exist $J\in U$ and $K\in W$ such that $B^{-1}_{JK}\neq 0$. For such a pair $(J,K)$ consider a contribution of terms from the expansions (\ref{vertex-series-high-degree})-(\ref{vertex-series-degree-1}) into (\ref{Zhat-plumbed-sl21-CT}) corresponding to fixed summation variables $\ell_I$. In the limit $\ell_J \gg \ell_K \gg \ell_S,\;\forall S\neq J,K$ it behaves as 
\begin{equation}
    \sim q^{\pm B_{JK}^{-1}\ell_J\ell_K}
\end{equation}
in the leading order. Moreover, the signs in the exponent are opposite for the terms coming from two different sums in the expansions $(a)$ or $(b)$. Therefore (\ref{Zhat-plumbed-sl21-CT}) in this chamber has necessarily arbitrary large positive and large negative powers of $q$. Therefore the set $U$ must actually be empty. 

Taking this into account, a good chamber must be necessarily of the form (\ref{chamber-alpha}), specified by some vector $\alpha\in \{\pm 1\}^{\vert|_{\deg\neq 2}}$. To argue the inequalities (\ref{plumbing-cond-1})-(\ref{plumbing-cond-3}) we will consider again the contributions of terms from the expansions (\ref{vertex-series-high-degree})-(\ref{vertex-series-degree-1}) into (\ref{Zhat-plumbed-sl21-CT}) corresponding to fixed summation variables $\ell_I$. 

In the regime when $\ell_J$ is very large for at least one $J\in \vert|_{\deg >2}$, compared to  $\ell_K,\;\;\forall K \in \vert|_{\deg = 1}$, the contribution behaves as
\begin{equation}
    \sim  q^{-\hspace{-1mm}\sum\limits_{I,J\in \vert|_{\deg>2}} \hspace{-2mm} \alpha_I\alpha_JB_{JK}^{-1}\ell_J\ell_K}
\end{equation}
and the positivity of the exponent is equivalent to the copositivity of the matrix $X_{IJ}=-B^{IJ}\alpha_I \alpha_J$. 

Fix $J\in \vert|_{\deg =1}$, $K\in \vert|_{\deg>2}$ and consider the regime $\ell_J \gg \ell_K \gg \ell_S,\;\forall S\neq J,K$. The contribution then behaves as
\begin{equation}
    \sim q^{-B_{JK}^{-1}\alpha_J\alpha_K\ell_J\ell_K}
\end{equation}
in the leading order. The non-negativity of the exponent is equivalent to the condition (\ref{plumbing-cond-2}). In case $B_{JK}^{-1}=0$, the subleading term in the exponent will dominate. It is positive due to copositivity of $X$.

Next fix a pair $J,K\in \vert|_{\deg =1}$, $J\neq K$ and consider the regime $\ell_J\sim \ell_K  \gg \ell_S,\;\forall S\neq J,K$. The contribution then behaves as
\begin{equation}
    \sim q^{-B_{JK}^{-1}\alpha_J\alpha_K\ell_J\ell_K}
\end{equation}
in the leading order. The positivity of the exponent is equivalent to the condition (\ref{plumbing-cond-3}).

Consider now a general regime of large $\ell\in \Z^L$ (i.e. at least one $\ell_I$ is large). The contribution behaves as
\begin{equation}
    \sim  q^{- \hspace{-1mm}\sum\limits_{J,K\in \vert|_{\deg>2}} \hspace{-2mm} \alpha_J\alpha_KB_{JK}^{-1}\ell_J\ell_K
   \; - \hspace{-1mm}\sum\limits_{\tiny\substack{J\in \vert|_{\deg>2}\\K\in \vert|_{\deg=1} }} \hspace{-2mm} \alpha_J\alpha_KB_{JK}^{-1}\ell_J\ell_K
\; - \hspace{-1mm}\sum\limits_{J,K\in \vert|_{\deg=1}} \hspace{-2mm} \alpha_J\alpha_KB_{JK}^{-1}\ell_J\ell_K
    }
\end{equation}
The exponent is then generically positive due to the conditions (\ref{plumbing-cond-1})-(\ref{plumbing-cond-3}). The only issue is the special direction when $\ell_I=0$ for all $I$ except some $I_0\in \vert|_{\deg=1}$. The exponent of $q$ is then identically zero for all such contributions from the expansions (\ref{vertex-series-high-degree})-(\ref{vertex-series-degree-1}). This can give an infinite number of non-trivial contributions to the constant term in $\widehat{Z}^{\mathfrak{sl}(2|1)}_{ab}$. However, one can make the sum of all such contributions to the constant term finite using the standard $\zeta$-function regularization, as described in the Section \ref{sec:Zhat-sl21} (which is essentially equivalent to the regularization by $\epsilon$ described in Appendix \ref{app:gauss-sums}).

\section{Asymptotic expansion of certain $q$-series}
\label{app:asympt-F}

Consider the following $q$-series:
\begin{equation}
	F(q;\alpha,\beta,A,B):=\sum_{m\geq 0} \frac{q^{\alpha m^2+\beta m}}{1-q^{Am+B}}
	\label{lambert-gen}
\end{equation}
Assume $A,\beta\geq 0$, $\alpha,B>0$. The asymptotic expansion of $F(q;\alpha,\beta,A,B)$ can be obtained via Euler-Maclaurin summation formula:
\begin{equation}
	F(q;\alpha,\beta,A,B) \approx \int_0^\infty f(x)dx -\sum_{r\geq 1}\frac{B_r^-}{r!}\,f^{(r-1)}(0)
	\label{EM}
\end{equation}
where
\begin{equation}
	f(x):=\frac{q^{\alpha x^2+\beta x}}{1-q^{Ax+B}}
\end{equation}
and we use the convention $B^\pm_1=\pm\frac{1}{2}$ for the two types of Bernoulli numbers\footnote{Namely $t/(e^{t}-1)=\sum_{m\geq 0} B^-_mt^m/m!$ and $t/(1-e^{-t})=\sum_{m\geq 0} B^+_mt^m/m!$.}. The expansion of $f(x)$ at $x=0$ reads
\begin{multline}
	f(x)=\sum_{M,N\geq 0}\frac{(-1)^N (\alpha x^2+\beta x)^N (Ax+B)^{M-1} B^+_{M}\hbar^{N+M-1}}{M!N!} =\\
	\sum_{N_1,N_2\geq 0}\sum_{M_1,M_2}\frac{(-1)^{N_1+N_2}\alpha^{N_1}\beta^{N_2}A^{M_1}B^{M_2}x^{2N_1+N_2+M_1}\,C_{M_1,M_2}\hbar^{N_1+N_2+M_1+M_2}}{N_1!N_2!}
\end{multline}
where
\begin{equation}
	C_{M_1,M_2}:=\left\{
	\begin{array}{ll}
		\frac{B^+_{M_1+M_2+1}}{M_1!M_2!(M_1+M_2+1)}, & M_1,M_2\geq 0, \\
		(-1)^{M_1}, & M_2=-M_1-1,\,M_1\geq 0, \\
		0, & \text{otherwise}. \\
	\end{array}
	\right.
\end{equation}
The second part of the (\ref{EM}) then reads:
\begin{multline}
	-\sum_{r\geq 1}\frac{B_r^-}{r!}\,f^{(r-1)}(0) = \\
	-\sum_{N_1,N_2\geq 0}\sum_{M_1,M_2}\frac{(-1)^{N_1+N_2}\alpha^{N_1}\beta^{N_2}A^{M_1}B^{M_2}\,B^-_{2N_1+N_2+M_1+1}\,C_{M_1,M_2}\hbar^{N_1+N_2+M_1+M_2}}{N_1!N_2!(2N_1+N_1+M_1+1)}\\
	= \sum_{L} {L!\tilde{\tilde{c}}_L}\hbar^L
\end{multline}
 where
\begin{equation}
	\tilde{\tilde{c}}_L=\sum_{\substack{N_1+N_2+M_1+M_2=L \\ N_{1,2}\geq 0}}\frac{(-1)^{N_1+N_2+1}\alpha^{N_1}\beta^{N_2}A^{M_1}B^{M_2}\,B^+_{2N_1+N_2+M_1+1}\,C_{M_1,M_2}}{L!N_1!N_2!(2N_1+N_1+M_1+1)}.
	\label{cL-expr}
\end{equation}
The contribution to the power low behavior in the asymptotics of $c_L$ when $L\rightarrow \infty$ will be given by the terms in the sum at the boundary of the summation region in (\ref{cL-expr}) where $M_1,N_2,M_2\ll L$. Using the formula
\begin{equation}
	B^\pm_{2n}=\frac{(-1)^{n+1}2(2n)!}{(2\pi)^{2n}}\,\zeta(2n),
\end{equation}
the fact the $B_r^+=0$ for odd $r\geq 1$, Stirling approximation
\begin{equation}
	n!=\sqrt{2\pi n}\left(\frac{n}{e}\right)^n(1+O(1/n)),
\end{equation}
and
\begin{equation}
    \zeta(2n)=\sum_{K\geq 0} \frac{1}{K^{2n}}
\end{equation}
we obtain that in $L\rightarrow\infty$ limit:
\begin{multline}
\tilde{\tilde{c}}_L=
\sum_{N_2,M_1,M_2}\frac{(-1)^{L-M_1-M_2+1}\alpha^{L-N_2-M_1-M_2}\beta^{N_2}A^{M_1}B^{M_2}\,C_{M_1,M_2}B^-_{2L-N_2-M_1-2M_2+1}}{L!(L-N_2-M_1-M_2)!N_2!(2L-N_2-2M_2+1)}
\\
\approx \sum_{N_2+M_1 \text{odd},M_1,K\geq 0}
\frac{i}{\pi K\sqrt{\pi L}}\,\left(\frac{\alpha}{K^2\pi^2}\right)^L
\times
\\
\frac{1}{N_2!}\left(\frac{i\pi K\beta}{\alpha}\right)^{N_2}
\left(\frac{\pi^2 K^2B}{\alpha L}\right)^{M_2}
\left(\frac{-i\pi K A}{\alpha}\right)^{M_1}
C_{M_1,M_2}(1+O(\frac{1}{L}))=\\
=\sum_{K\geq 0}\frac{-1}{\pi K \sqrt{\pi L}}\,\left(\frac{\alpha}{K^2\pi^2}\right)^L\,
\mathrm{Im}\,\frac{e^{\frac{\pi iK\beta}{\alpha}}}{1-e^{\frac{\pi i KA}{\alpha}}}\,(1+O(\frac{1}{L})).
\label{cL-as}
\end{multline}
We assumed that 
\begin{equation}
	K A/\alpha < 2
\end{equation}
so that the sum with resepect to $M_1$ above is convergent. This asymptotic behavior corresponds to the singularities in the Borel plane at $\xi=\frac{\pi^2\,K^2}{\alpha}$. From the general formula (\ref{stokes-jumps}) it follows that the  corresponding Stokes jumps are of the form
\begin{equation}
    \Delta Z = {-2 i}{\sqrt\frac{\pi}{\hbar\alpha}}  
	\,
	\mathrm{Im}\,\frac{e^{\frac{\pi iK\beta}{\alpha}}}{1-e^{\frac{\pi i KA}{\alpha}}}
	\,e^{-\frac{K^2\pi^2}{\hbar\alpha}}\,(1+O(\hbar)).
	\label{F-stokes-jump-type-I}
\end{equation}	
Now consider the asymptotic expansion of the first term in (\ref{EM}) at $\hbar\rightarrow \infty$:
\begin{multline}
	\int_0^\infty f(x)dx  =
	\frac{1}{\sqrt{\hbar\alpha}}
	\int_{0}^{+\infty} dt\frac{e^{-t^2-\beta\frac{\hbar}{\alpha}\,t}}{1-e^{-A\sqrt\frac{\hbar}{\alpha}\,t -B\hbar}} 
	\\
	\approx \frac{e^{\frac{\beta\hbar B^2}{A^2}}}{A\hbar} 
	F\left(\frac{B\sqrt{\hbar\alpha}}{A}\right)+
	\sum_{\substack{L\geq -1 \\ L\in \frac{1}{2}\mathbb{Z}}} \tilde{c}_L \Gamma\left(L+1\right)\hbar^{L }
\end{multline}
where 
\begin{multline}
    F(x):=\frac{e^{-x^2}}{2}
    (\pi\mathrm{erfi}(x)-\mathrm{Ei}(x))
    =\\ \sqrt\pi \sum_{n\geq 0}\frac{(-1)^n2^nx^{2n+1}}{(2n+1)!!}
-\frac{e^{-x^2}}{2}\left(\gamma+2\log x
+\sum_{n\geq 0}\frac{x^{2n}}{n\cdot n!}
\right)
\label{F-def}
\end{multline}
is a function which can be expressed as a series with infinite radius of convergence (therefore its Borel transform does not have singularities away from the origin) and 
\begin{equation}
    \tilde{c}_L=
    \frac{1}{2}\sum_{\frac{N+M_1}{2}+M_2-\frac{1}{2}=L}
    \frac{(-\beta)^N A^{M_1} B^{M_2}\alpha^{-\frac{N+M_1+1}{2}}C_{M_1,M_2}\,\Gamma\left(\frac{N+M_1+1}{2}\right)}{\Gamma(L+1)\,N!}.
    \label{cLprime-expr}
\end{equation}
At $L\rightarrow\infty$ we have
\begin{equation}
    \tilde{c}_L\approx \sum_{K\geq 0}
    \frac{A}{(2\pi K)^2 \alpha }
    \mathrm{Re}\left[
    \left(-\frac{A^2}{(2\pi K)^2\alpha}\right)^{L}\,
    e^{2\pi iK(\beta/A-2B\alpha/A^2)}(1+O(\frac{1}{L}))
    \right].
    \label{cLprime-as}
\end{equation}
It follows that the Borel transform has corresponding singularities at $\xi=-4\pi^2K^2\alpha/A^2$.
This is in agreement with the observation that the Stokes jumps in the first term of (\ref{EM}) originate from the poles in the integrand at $x=\pm\frac{2\pi i K}{A\hbar}-\frac{B}{A}$, $K\in \mathbb{Z}_+$. They are of the form
\begin{equation}
    \Delta Z =-\frac{2\pi i}{\hbar\,A}\,
	e^{\pm 2\pi iK(\beta/A-2B\alpha/A^2)}
	\, e^{\frac{(2K\alpha/A)^2\pi^2}{\hbar \alpha}}
	\,(1+O(\hbar)).
		\label{F-stokes-jump-type-II}
\end{equation}
The sign in the exponential depends on the choice of the branch of the Borel sum $B(\xi)$, which has a branching point at the origin.

To summarize, the asymptotic expansion of (\ref{lambert-gen}) is given by
\begin{equation}
    \sum_{m\geq 0} \frac{q^{\alpha m^2+\beta m}}{1-q^{Am+B}}
	\approx \frac{e^{\frac{\beta\hbar B^2}{A^2}}}{A\hbar} 
	F\left(\frac{B\sqrt{\hbar\alpha}}{A}\right)+
	\sum_{L\in \frac{1}{2}\mathbb{Z}_+} \tilde{c}_L \Gamma\left(L+1\right)\hbar^{L}+ \sum_{L\in \Z_+} {L!\tilde{\tilde{c}}_L}\hbar^L
\end{equation}
where the function $F$ is defined by a \textit{convergent} series in (\ref{F-def}) and the coefficients $\tilde{\tilde{c}}_L$ and $\tilde{c}_L$ are given by the explicit expressions (\ref{cL-expr}) and (\ref{cLprime-expr}) with asymptotic behavior at $L\rightarrow\infty $ as in (\ref{cL-as}) and (\ref{cLprime-as}).

\section{Quantum Modularity for $L(2,1)$}
\label{app:zeta}

The aim of this appendix is to provide two proofs of the transformation properties of $\widehat{Z}^{\mathfrak{sl}(2|1)}_{a,b}(\tau)$ for the lens space $L(2,1)$ reported in Section \ref{sec:q-mod-lens}. 

First, notice that these homological blocks can be expressed in terms of the Lambert series \eqref{eq:F} as
\begin{align}
\widehat{Z}^{\mathfrak{sl}(2|1)}_{0,0}(\tau)&= -\frac{1}{6}+2F(2\tau),\\
\widehat{Z}^{\mathfrak{sl}(2|1)}_{1,0}(\tau)&= \frac{1}{12} + 2F(\tau)-2F(2\tau),\\
\widehat{Z}^{\mathfrak{sl}(2|1)}_{1,1}(\tau)&= 2F(\tau/2)-4F(\tau)+2F(2\tau).
\end{align}
Then, equation \eqref{eq:q-mod-L21-T} immediately follows from the above relations and the invariance of $F(\tau)$ under $\tau\rightarrow \tau +1$.
Equation \eqref{eq:q-mod-L21-S} is readily derived from the relations above together with \eqref{eq:periodF} and  \eqref{eq:pF}.

Another proof of equation \eqref{eq:q-mod-L21-S}, along the lines of \cite{Bettin_2013, Lewis_2019}, is reported below.
The $\mathfrak{sl}(2|1)$ homological blocks of $L(2,1)$ can be written via the inverse Mellin transform\footnote{The Mellin transform of $f:(0,\infty)\rightarrow \mathbb{C}$ is
\begin{equation}
\mathcal{M}(f)(s):= \int_{0}^{\infty} t^{s-1} f(t) dt.
\end{equation}
Assume that the  the function $f(t)$ decays rapidly at  infinity and grows like $t^{-\alpha}$ at 0, then $\mathcal{M}(f)(s)$ is holomorphic on $\text{Re}(s)>\alpha$ domain. For $\sigma>\alpha$ and $t>0$ by the Mellin inversion formula we have
\begin{equation}
f(t)= \frac{1}{2\pi i}\int_{\text{Re}(s)=\sigma} t^{-s} \mathcal{M}(f)(s) ds.
\end{equation}
Extension of the above formula to functions with different growths can be found for instance in \cite{Zagier_Mellin}. 
The Mellin transform of the exponential function $f(t)=e^{-t}$ converges for $\text{Re}(s)>0$ and equals the gamma function $\Gamma(s)$.} as
\begin{align}
\widehat{Z}^{\mathfrak{sl}(2|1)}_{a,b}(\tau) 
&=  \text{const}_{a,b} + \frac{1}{\pi i}\int_{\text{Re}(s)=3/2} ds\, e^{\pi i s/2} (2\pi \tau)^{-s} \Gamma(s) 2^{-s}\zeta(s,a/2)\zeta(s,b/2)
\end{align}
where we assume that $|\text{arg}(-i\tau)|<\pi /2$ and use the fact that the Dirichlet series associated to $(\widehat{Z}^{\mathfrak{sl}(2|1)}_{a,b}(\tau)-\text{const}_{a,b})/2$ is 
\begin{equation}
L(d(a,b),s):=2^{-s}\sum_{k\ge 0}\sum_{n\ge 0} (k+b/2)^{-s}(n+c/2)^{-s}= 2^{-s}\zeta(s,a/2)\zeta(s,b/2)
\end{equation}
with $d(a,b)$ being an abbreviation for the coefficients defined in equation \eqref{Zhat-lens} in the case $p=2$ and 
\begin{equation}
\label{eq:hurwitz_zeta}
    \zeta(s,g):=\sum_{n\ge 0}\frac{1}{(n+g)^s}
\end{equation} being the Hurwitz zeta function, which reduces to the Riemann zeta function for $g=1$, and at $s=0$ is given by $\zeta(0,g)=-g+1/2$.

Moving the path of integration to $\text{Re}(s)=-1/2$, we obtain
\begin{align}
\widehat{Z}^{\mathfrak{sl}(2|1)}_{a,b}(\tau) 
&=  \text{const}_{a,b}+r_{2,(a,b)}(\tau)+\frac{1}{\pi i}\int_{\text{Re}(s)=-1/2} ds\, e^{\pi i s/2} (2\pi \tau)^{-s} \Gamma(s) L(d,s)\,.
\end{align}
The function $r_{2,(a,b)}(\tau)$ encodes the contributions from the poles of the Hurwitz zeta functions at $s=1$ and the simple pole at $s=0$ of the gamma function. Hence, we have
\begin{align}
r_{2,(a,b)}(\tau) &= \text{Res}_{s=1}\bigl((-2\pi i\tau)^{-s} 2\Gamma(s) L(d,s)\bigr) + \text{Res}_{s=0}\bigl((-2\pi i\tau)^{-s} 2\Gamma(s) L(d,s)\bigr) \notag\\
&= 
\frac{\text{log}(-4\pi i\tau)+\gamma-\gamma_0(a/2)-\gamma_0(b/2)}{4\pi i \tau} + \frac{1}{2}(a-1)(b-1).
\end{align}
Indeed, as $s$ approaches 1,
\begin{align}
&\zeta(s,g)=\frac{1}{s-1}+\sum_{n\ge 0}\frac{(-1)^n}{n!} \gamma_n(g)(s-1)^n,\\
&\Gamma(s) = 1-\gamma (s-1)+O(|(s-1)|^2),\\
&x^{-s} = x^{-1} - \frac{\text{log}(x)}{x}(s-1)+O(|(s-1)|^2),
\end{align}
where $\gamma_0(g)= -\psi(g)$ is the Digamma function and $\gamma_0(1)=\gamma$ the Euler-Mascheroni constant.

Consider now
\begin{align}
\frac{1}{2\tau} \bigl( &\widehat{Z}^{\mathfrak{sl}(2|1)}_{a',b'}(-1/\tau) -{\text{const}_{a',b'}}\bigr)=
\nonumber
\\
&=  \frac{1}{2\pi i\tau} \int_{\text{Re}(s)=3/2} ds\, e^{\pi i s/2} (2\pi)^{-s}(-1/\tau)^{-s} \Gamma(s) L(d(a',b'),s) \\
&= \frac{1}{2\pi i}\int_{\text{Re}(s)=3/2} ds\, e^{-\pi i s/2} (2\pi)^{-s}\tau^{s-1} \Gamma(s) 2^{-s} \zeta(s,a'/2)\zeta(s,b'/2) \\
&= \frac{1}{2\pi i}\int_{\text{Re}(s)=3/2} ds\, e^{-\pi i s/2} (2\pi \tau)^{s-1} \Gamma(1-s)2^{s-2} \frac{\text{sin}\bigl( \frac{s\pi}{2}\bigr)}{\text{cos}\bigl( \frac{s\pi}{2}\bigr)}\times\\
&\hspace{2cm}\times ((-1)^{a'}\zeta(1-s,1/2)+\zeta(1-s))((-1)^{b'}\zeta(1-s,1/2)+\zeta(1-s)) \notag\\
&= \frac{-i}{2\pi i}\int_{\text{Re}(s)=-1/2} ds\, e^{\pi i s/2} (2\pi \tau)^{-s} \Gamma(s) 2^{-s}\frac{\text{cos}\bigl( \frac{s\pi}{2}\bigr)}{\text{sin}\bigl( \frac{s\pi}{2}\bigr)}\times\\
&\hspace{2cm}\times 2^{-1}((-1)^{a'+b'}\zeta(s,1/2)^2+\zeta(s)^2+((-1)^{a'}+(-1)^{b'})\zeta(s,1/2)\zeta(s)).\notag
\end{align}
Above we assumed that $0<\text{arg}(\tau)<\pi$ and $0<\text{arg}(-1/\tau)<\pi$, and thus $\text{arg}(-1/\tau)=\pi-\text{arg}(\tau)$. To go from the third to the fourth line we used the functional equation of the Hurwitz zeta function 
\begin{equation}
\zeta(s,a/p)=2\Gamma(1-s)(2\pi p)^{s-1} \sum_{k=1}^p \text{sin}\biggl(\frac{\pi s}{2}+\frac{2\pi k a}{p}\biggr) \zeta(1-s,k/p)
\end{equation}
together with some identities satisfied by the gamma function. In the last step, we simply changed the integration variable from $s$ to $(1-s)$.

Finally, we obtain
\begin{align}
&\widehat{Z}^{\mathfrak{sl}(2|1)}_{a,b}(\tau) - \frac{1}{2\tau} \sum_{a',b'}(-1)^{aa'+bb'}\widehat{Z}^{\mathfrak{sl}(2|1)}_{a',b'}(-1/\tau) = \\
&= \widetilde{r}_{2,(a,b)}(\tau) + \frac{1}{\pi i}\int_{\text{Re}(s)=-1/2} ds\, e^{\pi is/2} (2\pi \tau)^{-s} \Gamma(s)L(d(a,b),s) \biggl(1+ i \frac{\text{cos}\bigl(\frac{s\pi}{2}\bigr)}{\text{sin}\bigl(\frac{s\pi}{2}\bigr)}\biggr)\\
&=\widetilde{r}_{2,(a,b)} + \frac{1}{\pi}\int_{\text{Re}(s)=-1/2} ds\, (2\pi \tau)^{-s}  \frac{\Gamma(s)}{\text{sin}\bigl(\frac{s\pi}{2}\bigr)}2^{-s}\zeta(s,a/2)\zeta(s,b/2) \label{eq:lastintegral}
\end{align}
where we denote by 
\begin{align}
\widetilde{r}_{2,(a,b)}(\tau) &= \text{const}_{a,b}+  {r}_{2,(a,b)}(\tau){-}\sum_{a',b'}(-1)^{aa'+bb'}\frac{\text{const}_{a',b'}}{2\tau} \notag\\
&= 
\text{const}_{a,b}+ \frac{1}{2}(a-1)(b-1)+\frac{1}{4\pi i \tau}{\times}\notag\\
&{\times} \bigl(\text{log}(-4\pi i\tau)+\gamma-\gamma_0(a/2)-\gamma_0(b/2){-}2\pi i\sum_{a',b'}(-1)^{aa'+bb'}\text{const}_{a',b'}\bigr)\,.
\end{align}
Note that the last integral \eqref{eq:lastintegral} converges for all $\tau\in \C$ with $|\text{arg}(\tau)|<\pi$ and thus it provides an extension of the period function $\psi_{2,(a,b)}(\tau)$ to the slit plane.

\section{Gauss sums and invariants for unrolled quantum groups}
\label{app:gauss-sums}

Consider the case when $M^3$ is obtained from a tree plumbing, that is $M^3$ is a surgery on the collection of linked unknots forming a tree. We will follow the conventions of Section \ref{sec:plumbed-3-manifolds}. Let us pick some node in the plumbing tree (corresponding to a particular unknot) and consider this tree as a rooted tree and with edges oriented according to the direction opposite to the root. Denote the index of the root node by $I_0$. The topological invariant of \cite{ha2018topological} corresponding to the quantum deformation parameter $\xi=q^\frac{1}{2}=\exp\frac{2\pi i}{\ell}$ (for odd $\ell\geq 3$) of  $\mathcal{U}_q(\mathfrak{sl}(2|1))$ then reads:
\begin{multline}
    N_\ell(M^3,\omega)=\\
    \sum_{s^I,t^I=0}^{\ell-1}d(\alpha^{I_0}_{s^{I_0}t^{I_0}})\prod_{I\in \text{Vert}} d(\alpha^I_{s^It^I})
    \langle \theta_{V_{\alpha^I_{s^It^I}}} \rangle^{B_{II}} 
    \prod_{(I,J)\in \text{Edges}} S'(\alpha^J_{s^Jt^J},\alpha^I_{s^It^I})
\end{multline}
where 
\begin{equation}
    \{x\}:=\xi^x-\xi^{-x},
\end{equation}
\begin{equation}
    \alpha_{st}:=(\alpha_1+s,\alpha_2+t)\;\in \mathbb{C}\times \mathbb{C},
\end{equation}
\begin{equation}
    (\alpha_1,\alpha_2)\equiv (\mu_1-\ell+1,\mu_2+\ell/2)\;\in \mathbb{C}\times \mathbb{C},
\end{equation}
\begin{equation}
    d(\alpha)=\frac{\{\alpha_1\}}{\ell\{\ell\alpha_1\}\{\alpha_2\}\{\alpha_1+\alpha_2\}},
\end{equation}
\begin{equation}
    S'(\alpha',\alpha)=\frac{1}{\ell d(\alpha)}
    \,\xi^{-4\alpha_2'\alpha_2-2\alpha_2'\alpha_1-2\alpha_1'\alpha_2},
\end{equation}
\begin{equation}
    \langle \theta_{V_\alpha} \rangle =
    -\xi^{-2\alpha_2^2-2\alpha_1\alpha_2}.
\end{equation}
The element $\omega\in H^1(M^3,\mathbb{Q}/\mathbb{Z}\times \mathbb{Q}/\mathbb{Z})\cong B^{-1}\Z/\Z\times B^{-1}\Z/\Z$ is specified by its values on the meridians $\mathfrak{m}_I,\; I\in \text{Vert}$ of the link components. That is
\begin{equation}
    \omega(\mathfrak{m}_I)=\mu^I=(\mu_1^I,\mu_2^I)\in \mathbb{Q}/\mathbb{Z}\times \mathbb{Q}/\mathbb{Z},\;\;\;\sum_{J}B_{IJ}\mu^J=0\mod \Z\times \Z.
\end{equation}
After some manipulations and change of summation variables $(a^I,b^I)=(s^I+t^I+1,t^I)$ we have:
\begin{multline}
    N_\ell(M^3,\omega)=
    \frac{\prod_{I\in\text{Vert}}(e^{2\pi i\mu_1^I}-e^{-2\pi i\mu_1^I})^{\text{deg}(I)-2}}{\ell^{L+1}}
    \times\\
    \times\sum_{a^I,b^I\in \Z_\ell}
F\left(\{\xi^{2(a^I+\mu_1^I+\mu_2^I)},\xi^{2(b^I+\mu_2^I)}\}_{I\in \text{Vert}}\right)\,
\xi^{-2\sum_{IJ}B_{IJ}(a^I+\mu_1^I+\mu_2^I)(b^J+\mu_2^J)}
\label{xi-invariant-massaged}
\end{multline}
where $F$ is the following rational function of $2L$ variables:
\begin{equation}
    F(y,z)=\prod_{I\in\text{Vert}}\left(
    \frac{y_I-z_I}{(1-y_I)(1-z_I)}
    \right)^{2-\text{deg}(I)}.
\end{equation}
We will need to use the following general Gauss reciprocity formula \cite{deloup2007reciprocity,jeffrey1992chern}:
\begin{multline}
\sum_{r\;\in\;\Z^N/\ell\Z^N}
\exp\left(\frac{2\pi i}{\ell}\,r^T\mathcal{M}r+\frac{2\pi i}{\ell}\,p^Tr\right)
=\\
\frac{e^{\frac{\pi i\sigma(\mathcal{M})}{4}}\,(\ell/2)^{N/2}}{|\det \mathcal{M}|^{1/2}}
\sum_{\delta\;\in\; \Z^N/2\mathcal{M}\Z^N}
\exp\left(-\frac{\pi i \ell}{2}(\delta+\frac{p}{\ell})^T\mathcal{M}^{-1}(\delta+\frac{p}{\ell})\right)
\label{reciprocity-alt}
\end{multline}
where $\mathcal{M}$ is a symmetric non-degenerate $N\times N$ matrix with integer entries and $\sigma(\mathcal{M})$ is its signature.

Consider then expansion of $F(y,z)$ with respect to $y_I,z_I$ in some chamber. In (\ref{xi-invariant-massaged}) we will need then to plug in   $y_I=\xi^{2(a^I+\mu_1^I+\mu_2^I)},\,z_I=\xi^{2(b^I+\mu_2^I)}$. Because $|\xi|=1$ this in principle violates the convergence of the series. However, one can cure it by introducing a regularization parameter $\epsilon>0$ so that the  arguments in $F$ are deformed to $y_I=e^{-\alpha_I\epsilon} \,\xi^{2(a^I+\mu_1^I+\mu_2^I)},\,z_I=e^{+\alpha_I\epsilon} \,\xi^{2(b^I+\mu_2^I)}$, where $\alpha_I(=\pm 1)$ determines the expansion chamber as in (\ref{chamber-alpha}). This makes the series to be convergent. Below we will consider analytic continuation with respect to $q=e^{\frac{4\pi i}{\ell}}$, away from the unit circle to the region with $|q|<1$. Taking the radial limit $q\rightarrow e^{\frac{4\pi i}{\ell}}$ and then $\epsilon\rightarrow 0$, we recover the original $N_\ell(M^3,\omega)$. On the other hand, one can first take the limit $\epsilon \rightarrow 0$. As we will see below, it will give a well defined $q$-series when the expansion chamber, specified by $\alpha$, is \textit{good}, in the terminology of Section \ref{sec:Zhat-sl21}. The result can be expressed using the $q$-series $\widehat{Z}^{\mathfrak{sl}(2|1)}_{b,c}$. The invariant $N_\ell(M^3,\omega)$ then can be recovered by taking the radial limit  $q\rightarrow e^{\frac{4\pi i}{\ell}}$, assuming that it commutes with the limit $\epsilon \rightarrow 0$. In our work we do not perform a rigorous mathematical analysis of whether the two limits commute but we will assume it.   

The contribution of the monomial $\prod_{I}y_I^{n_I}z_I^{m_I}$ (in the expansion of $F$ described above) to the sum in the second line of (\ref{xi-invariant-massaged}) reads
\begin{multline}
 \sum_{a^I,b^I\in \Z_\ell}
\xi^{-2\sum_{IJ}B_{IJ}(a^I+\mu_1^I+\mu_2^I)(b^J+\mu_2^J)+2\sum_I \left(n_I(a^I+\mu_1^I+\mu_2^I)+m_I(b^I+\mu_2^I)\right)}=\\
\frac{\,\ell^L}{2^L|\det B|}
\sum_{\tilde\beta,\tilde\gamma\in\Z^L/2B\Z^L}
e^{\pi i\ell \tilde\gamma^T B^{-1}\tilde\beta+2\pi i(n-B\mu_2)^TB^{-1}\tilde\gamma+2\pi i(m-M(\mu_1+\mu_2))^TB^{-1}\tilde\beta}
\cdot \xi^{2n^TB^{-1}m}=\\
\frac{\ell^L}{|\det B|}
\sum_{\beta,\gamma\in\Z^L/B\Z^L}
e^{2\pi i\ell \gamma^T B^{-1}\beta+4\pi i(n-B\mu_2)^TB^{-1}\gamma+2\pi i(m-B(\mu_1+\mu_2))^TB^{-1}\beta}
\cdot \xi^{2n^TB^{-1}m}.
\label{xi-invariant-reciprocity}
\end{multline}
To go from the first line to the second we used the Gauss reciprocity formula (\ref{reciprocity-alt}) with \begin{eqnarray}
    N & = & 2L \\
    r & = & \left( 
    \begin{array}{c}
       a   \\
        b
    \end{array}
    \right)  \\
    \mathcal{M} & = &
    -\left( 
    \begin{array}{cc}
       0  & B \\
        B & 0 
    \end{array}
    \right), \;\;\sigma(\mathcal{M})=\sigma(B)-\sigma(B)=0\\
      p & = &
    2\left( 
    \begin{array}{c}
       n-B\mu_2   \\
        m-B(\mu_1+\mu_2) 
    \end{array}
    \right) \\
    \delta & = &
    \left( 
    \begin{array}{c}
       \tilde\beta   \\
        \tilde\gamma
    \end{array}
    \right)  \\
\end{eqnarray}
From the second line to the third one, we made a change of variables $\tilde \beta= B\beta'+\beta,\;\beta'\in \Z_2^L,\,\beta\in \Z^L/B\Z^L$ and performed the sum over $\beta'$, which gives the $2^L$ factor times the delta function with the condition $\tilde\gamma =0\mod 2$, that is explicitly solved by $\tilde\gamma =2\gamma$.
Combining (\ref{xi-invariant-massaged}) and (\ref{xi-invariant-reciprocity}) we can write:
\begin{multline}
    N_\ell(M^3,\omega) =
   \frac{{\prod_{I\in\text{Vert}}(e^{2\pi i\mu_1^I}-e^{-2\pi i\mu_1^I})^{\text{deg}(I)-2}}}{\ell|\det B|}\times \\
   \times\sum_{\substack{\beta,\gamma \in \Z^L/B\Z^L \\ b,c\in B^{-1}\Z^L/\Z^L}}
e^{2\pi i\ell \gamma^T B^{-1}\beta+4\pi i(b-\mu_2)^T\gamma+2\pi i(c-(\mu_1+\mu_2))^T\beta}    \cdot (-1)^\Pi\, \widehat{Z}^{\mathfrak{sl}(2|1)}_{b,c}|_{q\rightarrow \xi^2}
\end{multline}
where, as in (\ref{Zhat-plumbed-sl21}),
\begin{multline}
	\widehat{Z}_{b,c}^{\mathfrak{sl}(2|1)}=(-1)^{\Pi}
	\mathrm{CT}_{y,z}\left\{
	\left.
	\left(\frac{y_I-z_I}{(1-z_I)(1-y_I)}\right)^{2-\deg(I)}
	\right|_{\text{chamber }\alpha}
	\right.
	\times\\
	\left.
	\sum_{\substack{n=Bb \mod B\Z^L \\ m=Bc \mod B\Z^L}}
	q^{\sum_{I,J}B^{-1}_{IJ}\,n_Im_J}\,
	\prod_{J} z_J^{m_J}y_J^{n_J} 
	\right\} \\ \;\; \in  q^{\Delta_{ab}}\Z[[q]].
	\label{Zhat-plumbed-sl21-CT-gauss}
\end{multline}
This relation has the following natural conjectural generalization for rational homology spheres:
\begin{multline}
    N_\ell(M^3,\omega) =
   \frac{\pm\mathcal{T}([2\omega_1])}{\ell|H_1(M^3,\Z)|}\times \\
   \times\sum_{\substack{\beta,\gamma \in H_1(M^3,\Z) \\ b,c\in H^1(M^3,\Q/\Z)}}
e^{2\pi i\ell \cdot \ell k(\beta,\gamma)+4\pi i(b-\omega_2)(\gamma)+2\pi i(c-(\omega_1+\omega_2))(\beta)}    \cdot \widehat{Z}^{\mathfrak{sl}(2|1)}_{b,c}|_{q\rightarrow \xi^2}
\end{multline}
where $\mathcal{T}$ is the Reidemeister torsion (equal to the analytic torsion) of the $U(1)$ flat connection $[2\omega_1]:=(2\omega_1\mod H^1(M^3,\Z))\in H^1(M^3,\Q/\Z)$, the same one as the one appeared in the context of $U(1|1)$ Chern-Simons theory \cite{ray1971r,muller1978analytic,cheeger1977analytic,Rozansky:1992zt,nicolaescu2008reidemeister,Mikhaylov:2015nsa}. The overall sign $\pm 1$ reflects the sign ambiguity in the definition of the torsion.

\bibliographystyle{JHEP}
\bibliography{sl2-1}

\end{document}